\documentclass[sigconf,nonacm]{acmart}
\AtBeginDocument{%
  }

\setcopyright{none}
\acmDOI{}
\acmISBN{}
\acmISBN{978-1-4503-XXXX-X/2018/06}

\usepackage{xspace}
\usepackage{listings}
\usepackage{xcolor}
\usepackage{subcaption}

\lstdefinestyle{prompt}{
  basicstyle=\ttfamily\footnotesize,
  breaklines=true,
  breakindent=0pt,
  columns=fullflexible,
  keepspaces=true,
  frame=single,
  rulecolor=\color{gray!40},
  backgroundcolor=\color{gray!5},
  xleftmargin=4pt,
  xrightmargin=4pt,
  aboveskip=6pt,
  belowskip=6pt,
}

\begin{document}

\title{GroundWork: Bridging Human and LLM Reasoning Through a Structured Argumentation Model}

\author{Sjoerd Vink}
\correspondingauthor
\authornotemark[1]
\email{s.a.vink@uu.nl}
\affiliation{%
  \institution{Utrecht University \& Tufts University}
  \city{Medford}
  \state{MA}
  \country{USA}
}

\author{Byungmoo Kim}
\email{byungmoo.kim@tufts.edu}
\affiliation{%
  \institution{Tufts University}
  \city{Medford}
  \state{MA}
  \country{USA}
}

\author{Michael Behrisch}
\affiliation{%
  \institution{Utrecht University}
  \city{Utrecht}
  \country{The Netherlands}
}

\author{Adam Coscia}
\affiliation{%
  \institution{Stevens Institute of Technology}
  \city{Hoboken}
  \state{NJ}
  \country{USA}
}

\author{Remco Chang}
\affiliation{%
  \institution{Tufts University}
  \city{Medford}
  \state{MA}
  \country{USA}
}

\renewcommand{\shortauthors}{Vink et al.}

\newcommand{\toolname}{GroundWork\xspace}

\newcommand{\sjoerd}[1]{\textcolor{orange}{(\textbf{Sjoerd:} #1)}}
\newcommand{\brian}[1]{\textcolor{orange}{(\textbf{Brian:} #1)}}
\newcommand{\remco}[1]{\textcolor{orange}{(\textbf{Remco:} #1)}}
\newcommand{\adam}[1]{\textcolor{orange}{(\textbf{Adam:} #1)}}
\newcommand{\mib}[1]{\textcolor{orange}{(\textbf{Michael:} #1)}}

\begin{abstract}
Large language models (LLMs) are increasingly integrated into workflows that require reasoning over text documents, yet interaction remains largely conversational.
This leaves argumentative structure implicit or embedded in text, making it difficult to maintain, compare, and revise arguments as analyses evolve.
We present \toolname, a mixed-initiative system that makes argument structure a shared, editable state for humans and LLMs.
\toolname operationalizes the Toulmin model of argumentation as a typed graph with explicit operations for human and LLM edits.
It exposes this structure through coordinated corpus views and an editable workbench for constructing, comparing, testing, and tracing source-grounded arguments.
We evaluate \toolname through an expert study with four participants spanning consulting, product development, law, and immunology, and three use-case scenarios in intelligence, law, and medicine.
Our findings show that GroundWork makes source-grounded arguments inspectable, comparable, and revisable as shared human–LLM state.
\end{abstract}

\begin{CCSXML}
<ccs2012>
 <concept>
  <concept_id>00000000.0000000.0000000</concept_id>
  <concept_desc>Do Not Use This Code, Generate the Correct Terms for Your Paper</concept_desc>
  <concept_significance>500</concept_significance>
 </concept>
 <concept>
  <concept_id>00000000.00000000.00000000</concept_id>
  <concept_desc>Do Not Use This Code, Generate the Correct Terms for Your Paper</concept_desc>
  <concept_significance>300</concept_significance>
 </concept>
 <concept>
  <concept_id>00000000.00000000.00000000</concept_id>
  <concept_desc>Do Not Use This Code, Generate the Correct Terms for Your Paper</concept_desc>
  <concept_significance>100</concept_significance>
 </concept>
 <concept>
  <concept_id>00000000.00000000.00000000</concept_id>
  <concept_desc>Do Not Use This Code, Generate the Correct Terms for Your Paper</concept_desc>
  <concept_significance>100</concept_significance>
 </concept>
</ccs2012>
\end{CCSXML}

\ccsdesc[500]{Do Not Use This Code~Generate the Correct Terms for Your Paper}
\ccsdesc[300]{Do Not Use This Code~Generate the Correct Terms for Your Paper}
\ccsdesc{Do Not Use This Code~Generate the Correct Terms for Your Paper}
\ccsdesc[100]{Do Not Use This Code~Generate the Correct Terms for Your Paper}

\begin{teaserfigure}
 \includegraphics[width=\textwidth]{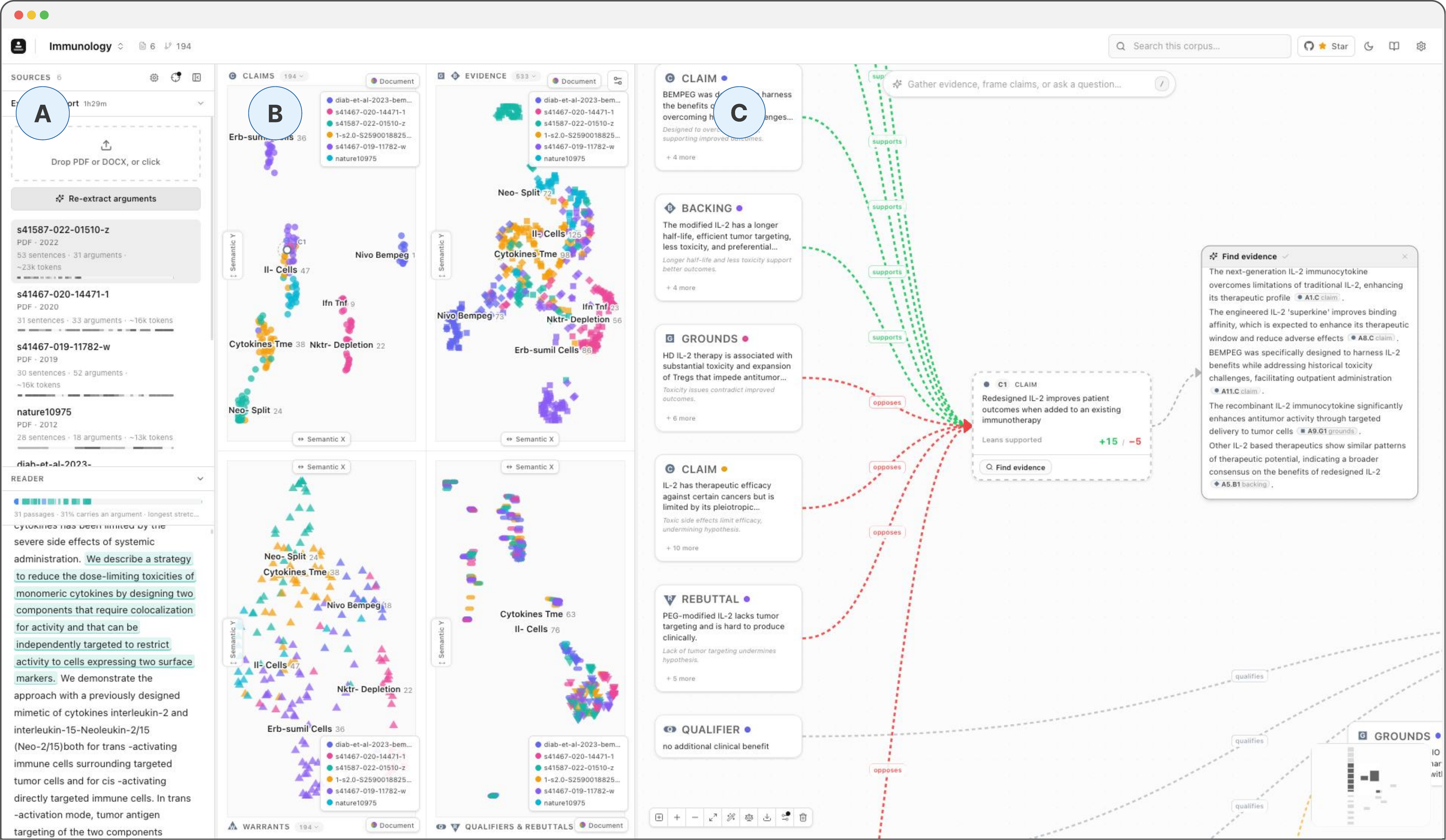}
  \caption{\textbf{The \toolname system enables collaborative argumentation for humans and LLMs.}
  The system decomposes arguments from uploaded documents into a structured argumentation schema.
  The corpus panel \textbf{(A)} organizes extracted argument components into four role-specific semantic spaces \textbf{(B)}.
  The workbench \textbf{(C)} provides an editable argument state where both humans and LLMs operate on.}
  \Description{Overview of GroundWork showing uploaded documents, role-specific views of extracted argument components, and an editable workbench where the user and LLM inspect and revise a shared argument state.}
  \label{fig:teaser}
\end{teaserfigure}

\received{20 February 2007}
\received[revised]{12 March 2009}
\received[accepted]{5 June 2009}

\maketitle

\section{Introduction}

Constructing and maintaining defensible arguments is a core task underlying many professionals' day-to-day workflows including law \cite{neumann2005legal}, medicine \cite{alonso2016grade}, and intelligence \cite{primer2009structured}.
Argumentation is inherently structured: experts must identify claims, connect them to evidence, examine the assumptions that justify those connections, and account for qualifiers and rebuttals \cite{toulmin2003uses}.
At the same time, document collections rarely present themselves in this structure.
The claims, evidence, assumptions, and counterarguments are often implicit and/or distributed across documents, requiring experts to continually reconstruct and maintain relationships.

To mitigate this burden, experts have increasingly turned to large language models (LLMs) due to their strong capabilities across argumentation tasks \cite{chen2024exploring, gao2024taxonomy}.
Prior HCI work has explored supporting experts with LLM-generated rationales \cite{wei2022chain,turpin2023language}, provenance interfaces \cite{martin2026papertrail}, argument maps \cite{zhang2023visar}, and argument mining \cite{li2025large}.
These systems and techniques have been applied to interpreting \cite{lam2024concept}, reviewing \cite{choe2024supporting}, and writing \cite{zhang2023visar} argumentative text, as well as making decisions based on such arguments \cite{ma2025towards}.
However, LLM interaction often remains organized through linear, text-heavy conversational interfaces, which provide limited cognitive support for organizing, comparing, and revising interdependent information across lengthy documents \cite{suh2023sensecape,Coscia:2025:OnGoal}.
Consider a lawyer using ChatGPT to assess a case: they may have the LLM reconcile statutes, prior rulings, expert reports, and newly introduced evidence on their behalf, but now have to manually track how each additional piece of information updates the case to develop an effective argument.
While good at search, LLMs by themselves are not designed to help users construct, evaluate, track, and revise arguments across multiple sources of text.
Further, when evidential relationships remain embedded in natural language, the experts and LLMs must repeatedly reconstruct how the argument fits together, risking the propagation of judgment errors.
Our goal is to bridge the gap between the structured nature of argumentation and text-based human-LLM interaction, where argument structure itself becomes explicit, shared, and interactive.

We first present a computational representation of structured argumentation for human-LLM interaction (Section~\ref{sec:intermediary-representation}), grounded in the Toulmin model of argumentation~\cite{toulmin2003uses}.
The functional roles of Toulmin (claim, grounds, warrant, backing, qualifier, rebuttal) correspond to recurring components of expert argumentation workflows.
Second, we formalize argumentation as a typed graph and introduce an operation vocabulary that specifies how humans and LLMs can construct, assess, and revise its components and relationships.
In doing so, our model brings the claims, evidence, assumptions, and counterarguments distributed across documents into a persistent shared structure that can be organized, compared, and revised throughout the analysis.
We instantiate this structured argumentation model in \toolname, a mixed-initiative system for reasoning over document corpora (Section~\ref{chap:system}).
\toolname organizes extracted argument components into coordinated corpus views and an interactive workbench where domain experts can construct arguments, compare claims, question evidence and assumptions, and trace components to their sources.
The LLM operates on the same data foundation to retrieve evidence, propose argument components, and support comparison, while domain experts control which proposals enter the agreed argument state.
Together, these interactions move argumentative relationships out of dispersed documents and conversational text into an explicit, inspectable, and revisable structure shared by both humans and LLMs.

We evaluate \toolname through three use-case scenarios and an expert study with four participants.
The use-cases demonstrate how \toolname can be applied in practice, supporting argumentation tasks in intelligence, law, and medicine.
The expert study revealed that similar argument structures can underlie substantially different reasoning processes and divisions of labor between humans and AI.
Participants distinguished the role evidence plays in an argument from its evidential standing, which depended on factors such as authority, credibility, recency, and context.
Finally, making reasoning explicit gives humans and LLMs identifiable objects to operate on, while also exposing tensions between fixed structure and the flexibility of expert practice.
Together, the findings show the promise of structured argumentation as an intermediary representation for externalizing, comparing, and revising argumentation, while also revealing important design trade-offs for future human-LLM systems.

In summary, our paper contributes:
\begin{itemize}
    \item A structured argumentation model, formalized as a typed Toulmin graph and a closed operation vocabulary.
    \item \toolname, an open-source system for human-LLM argumentation.
    \item An evaluation through three use-cases and an expert study, showing where structured argumentation matches how professionals reason, and where it requires domain-specific extension.
    \item Four design implications for building future HCI tools that better support interactive reasoning between humans and computational engines like LLMs.
\end{itemize}

\section{Related Work} \label{chap:related-work}

\subsection{Argumentation Models and Interactive Argument Mapping}

Argumentation research studies how claims are supported, challenged, and evaluated through structured relations between evidence, inference, and counterargument \cite{bench2007argumentation}.
It offers several models for representing and evaluating arguments.
Argumentation schemes characterize recurring patterns of defeasible reasoning and the critical questions used to evaluate them \cite{walton2002argumentation}.
Abstract argumentation frameworks model arguments and attack relations in order to determine which arguments are acceptable \cite{dung1995acceptability}.
Pragma-dialectics treats argumentation as a dialogic process in which standpoints are defended against challenge \cite{grootendorst2004systematic}.
The Argument Interchange Format further demonstrates how argument structures can be represented computationally across different argumentation formalisms \cite{chesnevar2006towards}.

HCI and argumentation systems have also long made argumentative relationships explicit and interactive.
For example, gIBIS represents issues, positions, and arguments as typed networks for collaborative deliberation \cite{conklin1988gibis}, while Araucaria supports the reconstruction and visual analysis of arguments using multiple argumentation schemes and notations \cite{reed2004araucaria}.
This work establishes an important precedent for treating arguments as computational objects that users can inspect and manipulate.

We build specifically on the Toulmin model because its functional roles provide interpretable reasoning objects for evidence-based work \cite{toulmin2003uses}.
Our focus is not on introducing computational argument representation itself, but on using such a representation as shared state between humans and LLMs reasoning over document corpora.

\subsection{LLMs for Argument Mining and Construction}

Recent work uses LLMs for various related tasks in computational argumentation, including extracting argument structure from text and generating new argumentative content \cite{chen2024exploring}.
In argument mining, LLMs can identify claims, premises, and support or attack relations with less task-specific modeling than earlier pipeline approaches \cite{li2025large}.
They also expand the scope of extraction beyond explicit statements by supporting the reconstruction of implicit or unstated argumentative content \cite{sviridova2026mining}.
This is important for our setting because warrants, assumptions, and rebuttals are often not stated directly in source documents.

LLMs have also been used to construct argumentative material.
Prior work shows that models can generate arguments conditioned on a topic, stance, or aspect \cite{schiller2021aspect}, and can produce counterarguments against an existing conclusion \cite{alshomary2023conclusion}.
Recent work further explores the Toulmin model as a schema for zero-shot argument explication, showing that LLMs can organize argumentative text into typed components such as claims, grounds, warrants, and rebuttals \cite{gupta2024harnessing}.
This work shows that LLMs can both extract and generate argumentative structure.
However, extraction and generation alone do not provide a complete interaction model.
Users still need to inspect, correct, compare, and revise the resulting structure.
This motivates our focus on a shared intermediary representation that supports mixed-initiative interaction, where users retain control over the reasoning process while the model helps construct and transform argument components \cite{horvitz1999mixedinitiative,amershi2019guidelines,heer2019agency}.

\subsection{Systems for Human-AI Argumentation and Sensemaking}

Prior visual analytics work has used interactive visual representations to support analysts in organizing, relating, and revisiting information during analysis.
Jigsaw supports investigative sensemaking across text collections through coordinated views of documents, entities, and their relationships \cite{stasko2008jigsaw}.
Semantic interaction goes further by allowing analysts to manipulate meaningful visual objects while those interactions steer underlying computational models \cite{endert2012semantic}.
More recently, rTisane showed how externalizing an analyst's conceptual model as an editable graph can surface assumptions and connect domain reasoning to subsequent computation \cite{jun2024rtisane}.
Together, these systems establish a broader precedent for coupling human reasoning and computation through persistent intermediate representations.

Recent human-LLM systems increasingly move beyond a linear transcript by externalizing parts of the interaction.
Graphologue converts LLM responses into interactive node-link diagrams \cite{jiang2023graphologue}, while Sensecape organizes LLM-supported exploration into hierarchical structures at multiple levels of abstraction \cite{suh2023sensecape}.
OnGoal externalizes conversational goals and their progression across long interactions \cite{Coscia:2025:OnGoal}.
These systems demonstrate how structure can make otherwise transient aspects of LLM interaction persistent and directly inspectable.

A closely related line of work structures sensemaking over document collections.
Synergi combines user-provided research threads, citation structure, and LLM-generated organization into an editable workspace for scholarly synthesis \cite{kang2023synergi}.
Marco uses user-defined schemas to support mixed-initiative extraction and comparison across business documents \cite{fok2024marco}, while LLooM lets analysts generate and refine high-level concepts over unstructured text \cite{lam2024concept}.
NEEDLE is particularly relevant to our setting because it treats claims as units of retrieval and consolidation, combining hypothesis-driven search, consistency checking, provenance, and a canvas for organizing evidence across a corpus \cite{duck2025finding}.

Other systems make structure explicit around writing and verification.
VISAR represents argumentative writing plans as editable hierarchical structures that users can revise with AI assistance \cite{zhang2023visar}.
PaperTrail decomposes LLM answers and source documents into claims and evidence, making claim-level provenance available for verification \cite{martin2026papertrail}.
RELIC similarly supports inspection of potentially unsupported content by comparing claims across sampled model responses \cite{cheng2024relic}.

These systems differ primarily in what they make explicit and persistent.
Prior work structures model responses, conversational goals, concepts, document schemas, research threads, writing plans, narratives, or claim-evidence provenance.
\toolname instead makes the evolving argument itself the shared computational state.
Claims, grounds, warrants, backing, qualifiers, and rebuttals remain source-grounded graph objects that humans and LLMs can inspect and transform throughout an analysis.
The distinction is therefore not between structured and unstructured interfaces, but between structure around an LLM-supported workflow and argument structure as the state through which human-LLM reasoning proceeds.
\begin{figure*}[t]
  \centering
  \includegraphics[width=\linewidth]{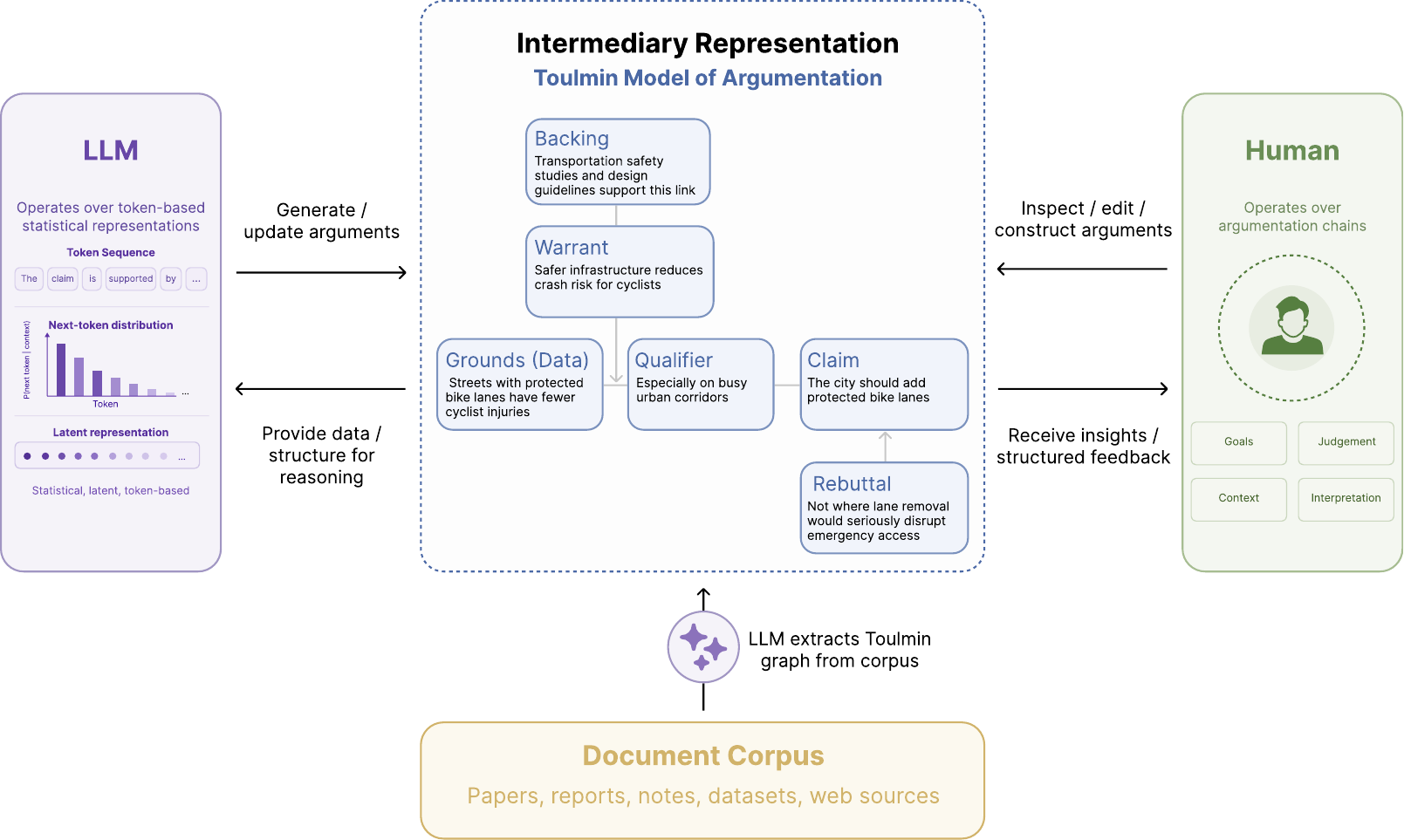}
  \caption{\textbf{The Toulmin model as an intermediary representation connecting human reasoning, AI reasoning, and document corpus}. The shared structure enables each party to inspect, construct, and revise arguments bidirectionally.}
  \Description{Conceptual diagram showing the Toulmin argument model as a shared intermediary representation between a human, an LLM, and a document corpus. Claims, grounds, warrants, backing, qualifiers, and rebuttals form the common structure through which arguments can be inspected, constructed, and revised.}
  \label{fig:ir_overview}
\end{figure*}

\section{Background: The Toulmin Model of Argumentation} \label{sec:toulmin}

To ground our intermediary representation of argumentation, we started from the Toulmin model of argumentation \cite{toulmin2003uses}, a widely used framework for analyzing argumentation. 
The model identifies the functional roles through which a claim is supported, qualified, and challenged, rather than asking whether an argument fits a fixed logical form.
This makes it useful for practical reasoning in domains where evidence is incomplete, assumptions matter, and conclusions are rarely absolute.

Consider a city debating protected bike lanes.
The model decomposes that argument into six roles:
\begin{itemize}
    \item A \emph{claim} is the assertion at stake: the city should add protected bike lanes.
    \item \emph{Grounds} are the evidence offered for that claim: streets with protected bike lanes have fewer cyclist injuries.
    \item A \emph{warrant} is the (often unstated) rule that makes the evidence relevant: safer infrastructure reduces crash risk for cyclists.
    \item A \emph{backing} supports that rule, not the claim itself: transportation safety studies and design guidelines.
    \item A \emph{qualifier} limits the strength or scope of the claim: especially on busy urban corridors.
    \item A \emph{rebuttal} states when the claim may not hold: not where removing a traffic lane would disrupt emergency access.
\end{itemize}

The example illustrates how Toulmin separates functions within an argument that may otherwise appear together in text.
Grounds provide evidence for a claim, warrants explain why that evidence is relevant, backing supports the warrant, and qualifiers and rebuttals capture where the claim is limited or challenged.
This decomposition makes disagreement more specific: a user may accept the evidence while questioning the warrant, narrow the claim through a qualifier, or introduce a rebuttal without rejecting the argument as a whole.

For our purposes, these roles provide a small set of reasoning objects that can be represented explicitly and acted on independently.
This makes Toulmin a useful foundation for an intermediary representation in which humans and LLMs can refer to, inspect, and revise the same parts of an argument as an analysis develops.
The next section operationalizes these roles and their relationships as a computational state shared by humans and LLMs.

\section{Structured Argumentation as Intermediary Representation} \label{sec:intermediary-representation}

The Toulmin model \cite{toulmin2003uses} provides a conceptual vocabulary for describing the components of an argument, but does not define how those components can serve as an interactive computational state.
We use these roles to define an intermediary representation in which argument components extracted from the document corpus are represented explicitly and can be referenced by both human and LLM actions (Figure~\ref{fig:ir_overview}).
We define the representation in two parts.
First, a typed argument graph with two layers specifying the argument components and the admissible relations between them.
Second, a closed operation vocabulary specifying how the graph can be logically transformed and edited.

\begin{table*}[t]
    \centering
    \small
    \begin{tabular}{llll}
    \toprule
    \textbf{Edge type} & \textbf{Layer} & \textbf{Typing} & \textbf{Reading} \\
    \midrule
    \multicolumn{4}{@{}l}{\emph{Structural} - within one argument} \\
    \texttt{grounds\_for} & Both & $G \rightarrow C$ & evidence offered for a claim \\
    \texttt{warrants} & Both & $W \rightarrow C$ & warrant for the grounds-claim step \\
    \texttt{backs} & Both & $B \rightarrow W$ & support for the warrant \\
    \texttt{qualifies} & Both & $Q \rightarrow C$ & scope or strength of the claim \\
    \texttt{rebuts} & Both & $R \rightarrow C \mid G \mid W$ & challenge to a claim, evidence, or inference \\
    \addlinespace

    \multicolumn{4}{@{}l}{\emph{Relational} - across arguments} \\
    \texttt{corroborates} & Both & admissible pairs & nodes that agree or mutually reinforce \\
    \texttt{replicates} & Corpus & $C \rightarrow C$ & claims that report the same finding \\
    \texttt{refines} & Corpus & $C \rightarrow C$ & claim that narrows or extends another \\
    \texttt{contends} & Corpus & $C \rightarrow C$ & claims in tension \\
    \texttt{similar} & Corpus & $C \rightarrow C,\; W \rightarrow W$ & shared semantic or inferential pattern \\
    \texttt{draws\_on} & Inquiry & admissible pairs & one component relies on another \\
    \texttt{undermines} & Inquiry & admissible pairs & one component weakens another \\
    \addlinespace

    \multicolumn{4}{@{}l}{\emph{Thematic} - analyst-defined grouping} \\
    \texttt{member\_of} & Inquiry & $v \rightarrow \Theta$ & theme membership \\
    \texttt{grounds\_for} & Inquiry & $\Theta \rightarrow C$ & grouped grounds for a claim \\
    \addlinespace
    
    \multicolumn{4}{@{}l}{\emph{Stance} - computed claim-comparison records} \\
    \texttt{supports} & Inquiry & $v \rightarrow C$ & item counts in favor of the claim \\
    \texttt{attacks} & Inquiry & $v \rightarrow C$ & item counts against the claim \\
    \bottomrule
    \end{tabular}
    \caption{\textbf{Edge vocabulary for the argument graph.}
    The layer column specifies whether an edge type can appear in the corpus graph, the inquiry graph, or both.
    Structural edges operationalize Toulmin's argument roles.
    Relational and thematic edges support corpus organization and analyst interpretation.
    The typed graph fixes the complete set of admissible endpoint pairs.}
    \label{tab:edge-vocabulary}
\end{table*}

\subsection{The Argument Graph}

Arguments are represented as a typed graph with two layers: the corpus graph and the inquiry graph.
The \emph{corpus graph} contains argument nodes and edges extracted from the document corpus.
It is immutable during the analysis and serves as a stable, source-grounded basis for inquiries.
The \emph{inquiry graph} contains the state of a single step in the analysis.
Its nodes either reference corpus nodes or are authored during the analysis;  
% Its edges record the argument structure being constructed, revised, or assessed.
its edges record the working argument as it is constructed, revised, or assessed.
This split matters for three reasons.
Analyses can be repeated and compared because the corpus stays fixed;
several inquiries can diverge over the same corpus;
and extraction errors are judged in the inquiry rather than written back into the corpus, so the extracted record is preserved.

Nodes take one of seven types:
claim ($C$), grounds ($G$), warrant ($W$), backing ($B$), qualifier ($Q$), rebuttal ($R$), and theme ($\Theta$).
The first six correspond to Toulmin-specific nodes.
A theme is a named group of nodes that records an analyst's organization of related information.
Each node stores its text, type, origin, provenance, and assessment state.
The origin records whether the component was extracted from the corpus or added later.
The provenance records the source or graph context from which the component was derived.
The assessment state records judgments such as confirmed, questioned, or rejected.

Edges specify how these components participate in an argument.
Table~\ref{tab:edge-vocabulary} details the edge vocabulary.
For each relation, the table identifies the graph layer in which it may occur, the admissible source and target node types, and its intended reading in the argument.
Structural edge types operationalize dependencies implied by Toulmin's roles.
The remaining edge types support corpus organization, analyst grouping, and claim comparison within the argument analysis and are not intended as a general ontology of argumentative relations.
An edge records a relation; it does not by itself justify a claim.
Instead, we represent a supported claim by an argument structure in which grounds provide evidence, a warrant licenses the inference from grounds to claim, backing supports the warrant, qualifiers constrain the claim, and rebuttals identify conditions under which the claim may fail.

\begin{table*}[t]
    \centering
    \begin{tabular}{llll}
    \toprule
    \textbf{Operation} & \textbf{Actor} & \textbf{Precondition} & \textbf{Effect} \\
    \midrule
    \texttt{MATERIALIZE(x)} & Human & $x$ is a corpus node & Add reference node to inquiry graph \\
    \texttt{AUTHOR(k,text)} & Human & $k$ is a valid node type & Add authored node of type $k$ \\
    \texttt{LINK(u,v,r)} & Human & $r$ is admissible for $u$ and $v$ & Add edge of type $r$ from $u$ to $v$ \\
    \texttt{RELABEL(e,r)} & Human & $e$ is editable, $r$ is admissible & Change edge $e$ to type $r$ \\
    \texttt{DELETE(x)} & Human & $x$ is editable & Remove node or edge $x$ \\
    \texttt{GROUP(S,label)} & Human & $S$ is a set of nodes, $|S| \geq 2$ & Add grouping over selected nodes \\
    \texttt{INTERPRET($\theta$,text)} & Human & $\theta$ is a theme & Add interpretation text to theme $\theta$ \\
    \texttt{DISSOLVE($\theta$)} & Human & $\theta$ is a theme & Remove theme grouping \\
    \texttt{ASSESS(x,j)} & Human & $x$ is a node, $j$ is an assessment label & Attach assessment $j$ to node $x$ \\
    \texttt{OVERRIDE(s,a)} & Human & $s$ is a stance item & Override model-assigned stance \\
    \texttt{SET\_ASIDE(s)} & Human & $s$ is a stance item & Temporarily neutralize stance item \\
    \texttt{RECORD(y)} & LLM & $y$ is a valid generated result & Record generated result \\
    \texttt{PROMOTE($\delta$)} & Human & $\delta$ is an accepted candidate edit & Apply candidate edit to inquiry graph \\
    \bottomrule
    \end{tabular}
    \caption{\textbf{Core operations over the inquiry graph.}
    Human operations directly transform the inquiry graph or its assessment state.
    LLM operations record generated results.
    Candidate graph edits are stored as generated results and enter the inquiry graph only through \texttt{PROMOTE}.}
    \label{tab:argument-operations}
\end{table*}

\subsection{State Changes in the Inquiry Graph}

The inquiry graph evolves through a closed vocabulary of operations, shown in Table~\ref{tab:argument-operations}.
Each row specifies an operation, who may perform it, the condition under which it is valid, and the resulting change to the inquiry state.
Operations whose preconditions are not satisfied leave the graph unchanged.
We include actors in the formalism because \toolname's aim is to develop a mixed-initiative system (see Section~\ref{chap:system}).
% The representation is shared by the human and the LLM, but they do not have the same authority over it.
% Humans may edit the inquiry graph directly, while an LLM can only record generated results, including so-called candidate graph edits.
% Candidate edits do not change the inquiry graph until they are accepted by a human through the \texttt{PROMOTE} operation.
% This preserves human agency and control over the reasoning state while allowing the LLM to help construct and revise arguments.
The representation is shared by the human and the LLM, but they do not have equal authority over it.
Humans can edit the inquiry graph directly; the LLM can only record generated results, including candidate graph edits.
A candidate edit does not change the inquiry graph until a human accepts it through the \texttt{PROMOTE} operation.
This keeps control of the reasoning state with the human, while still allowing the LLM to help construct and revise arguments.
Appendix \ref{app:example_toulmin} provides a working example illustrating how these state changes affect an argument.

% Section~\ref{chap:system} describes how these abstract operations are exposed through the interface.

% Derived judgments are computed from the current inquiry graph.
% For example, whether a claim is well supported depends on the active Toulmin structure around the claim, the stance edges produced during comparison, and any human assessments or overrides.
% When a human edits, deletes, or assesses a node or edge, the judgment should then be recomputed from the updated graph state.

% The formalism defines what a valid argument state is and how it can change.
% Section \ref{chap:system} describes how \toolname operationalizes this representation: first by using an LLM to extract an initial corpus graph, and then by providing an interface through which humans and the LLM inspect, revise, and extend an inquiry graph.
\section{Design Requirements for \toolname} \label{sec:design_process}

We derive the design requirements for \toolname from three complementary sources.
Argumentation theory identifies the components and relationships that must be represented \cite{toulmin2003uses}.
Domain literature in intelligence, medicine, and law describes how experts construct, compare, challenge, and evaluate arguments in practice \cite{primer2009structured,heuer1999psychology,alonso2016grade,neumann2005legal}.
Prior HCI work provides interaction principles for externalizing reasoning, navigating document collections, maintaining provenance, and coordinating human and model contributions (see Section \ref{chap:related-work}).
% \remco{need citation, or say something like, see related work section}.

Across these sources, two recurring problems motivate the system design.
Argument components are often distributed across documents or left implicit, while their relationships must be reconstructed during analysis.
Conversational LLM interfaces provide limited support for maintaining and revising these relationships as the analysis changes.
We translate these problems into five interaction requirements for \toolname:

\begin{itemize}
    \item[\textbf{R1}] \textbf{Explore the argument structure of the document corpus}.
    Analysts should be able to identify corpus-level patterns and move between related claims, evidence, local argument structures, and their source passages.
    This follows from sensemaking and visual analytics work on moving between overview, detail, and externalized structure \cite{pirolli2005sensemaking,zhang1994representations,hollan2000distributed}, and extends LLM interfaces for document synthesis and provenance toward argument-structure navigation \cite{martin2026papertrail,suh2023sensecape}.

    \item[\textbf{R2}] \textbf{Curate claims and rebuttals relevant to arguments}.
    Analysts should be able to construct arguments from extracted, authored, and generated claims, evidence, assumptions, qualifiers, and rebuttals, while controlling which model-generated candidates enter the argument state.
    This reflects analytic practices of generating hypotheses, alternatives, objections, and evidence \cite{primer2009structured,bardach2023practical}, as well as mixed-initiative guidance emphasizing inspectable and controllable model suggestions \cite{horvitz1999mixedinitiative,amershi2019guidelines}.
    
    \item[\textbf{R3}] \textbf{Compare multiple claims and evidence at once}.
    Analysts should be able to compare claims alongside the evidence, assumptions, qualifiers, and rebuttals that affect their support, exposing why judgments differ and what could change them.
    This follows from intelligence and policy methods for evaluating alternatives and tradeoffs, and medical frameworks that distinguish evidence certainty from other judgment criteria \cite{primer2009structured,heuer1999psychology,bardach2023practical,alonso2016grade}.

    \item[\textbf{R4}] \textbf{Test argument logic interactively}.
    Reasoning should remain editable, allowing analysts to modify, assess, or remove argument components and observe how those changes affect the current argument state.
    This builds on work on external representations and epistemic action \cite{kirsh1994epistemic,kirsh2010thinking,zhang1994representations}, and on human-AI interaction research emphasizing the ability to inspect, correct, and calibrate model outputs \cite{amershi2019guidelines,bucinca2021trust,sivaraman2023ignore,cheng2024relic}.

    \item[\textbf{R5}] \textbf{Trace evidence back to source documents}.
    Analysts should be able to inspect the source context behind claims and evidence, including the assumptions and inferential links connecting evidence to conclusions.
    This supports verification of grounded claims \cite{martin2026papertrail} and reflects domain practices where conclusions depend on the quality, relevance, and limitations of their evidence \cite{primer2009structured,alonso2016grade}.
\end{itemize}

\section{The Groundwork System} \label{chap:system}

\toolname is a visual analytics system for human-LLM reasoning over document corpora.
It instantiates the intermediary representation from Section~\ref{sec:intermediary-representation} as a persistent argument state that both humans and LLMs operate on.
An extraction pipeline first constructs the corpus graph from the uploaded documents, after which the interface exposes this structure through coordinated corpus views and an editable workbench.
We describe the extraction pipeline in Appendix~\ref{sec:argument-graph-extraction} and treat the resulting corpus graph as given, focusing on how humans and LLMs inspect and revise that argument structure throughout an analysis.

\subsection{System Overview}
\label{sec:system-overview}

The \toolname interface consists of three coordinated views: a corpus panel for source tracing (\textbf{R5}), a corpus-level argument view for exploration (\textbf{R1}), and a workbench for constructing, comparing, and revising arguments (\textbf{R2}-\textbf{R4}).

The \emph{corpus panel} (A in Figure~\ref{fig:teaser}) shows the uploaded documents and links argument components with their source.
A navigation bar indicates where extracted arguments occur throughout the document.
Together, these elements preserve the connection between the argument representation and its original document.

The \emph{argument view} (B in Figure~\ref{fig:teaser}) exposes the extracted argument graph.
Components are separated into four coordinated spaces: claims, grounds and backing, warrants, and qualifiers and rebuttals.
This organization builds on semantic substrates \cite{shneiderman2006network}, which place nodes into meaningful regions before examining relations across them.
Within each space, component embeddings are projected to reveal semantic clusters within that argumentative role.
Analysts can change the axes to organize nodes by attributes such as topic, source document, or node type.
Semantic zooming adds TF-IDF topic labels to local regions as analysts inspect dense clusters.
Selecting a node or brushing over a region highlights related components across the other spaces and in the source panel.
Hovering over a brushed region for two seconds opens a summary of the enclosed nodes.

The \emph{workbench} (C in Figure~\ref{fig:teaser}) exposes the editable inquiry graph as the current argument state.
Analysts can add or author nodes, revise relations, group components into themes, and record assessments, implementing the operation vocabulary from Section~\ref{sec:intermediary-representation}.
A natural-language input above the workbench provides an additional way to act on this state.
Analysts can request evidence, candidate claims, rebuttals, missing warrants, or claims to test.
The LLM model returns structured graph content rather than free text, such as candidate nodes, related evidence, or comparison results.
This preserves a flexible natural-language entry point while keeping model proposals distinct from human commitment.

\begin{figure*}[!t]
    \centering

    \begin{minipage}[t]{0.43\linewidth}
        \vspace{0pt}
        \begin{subfigure}[t]{\linewidth}
            \centering
            \includegraphics[
                width=\linewidth,
                keepaspectratio
            ]{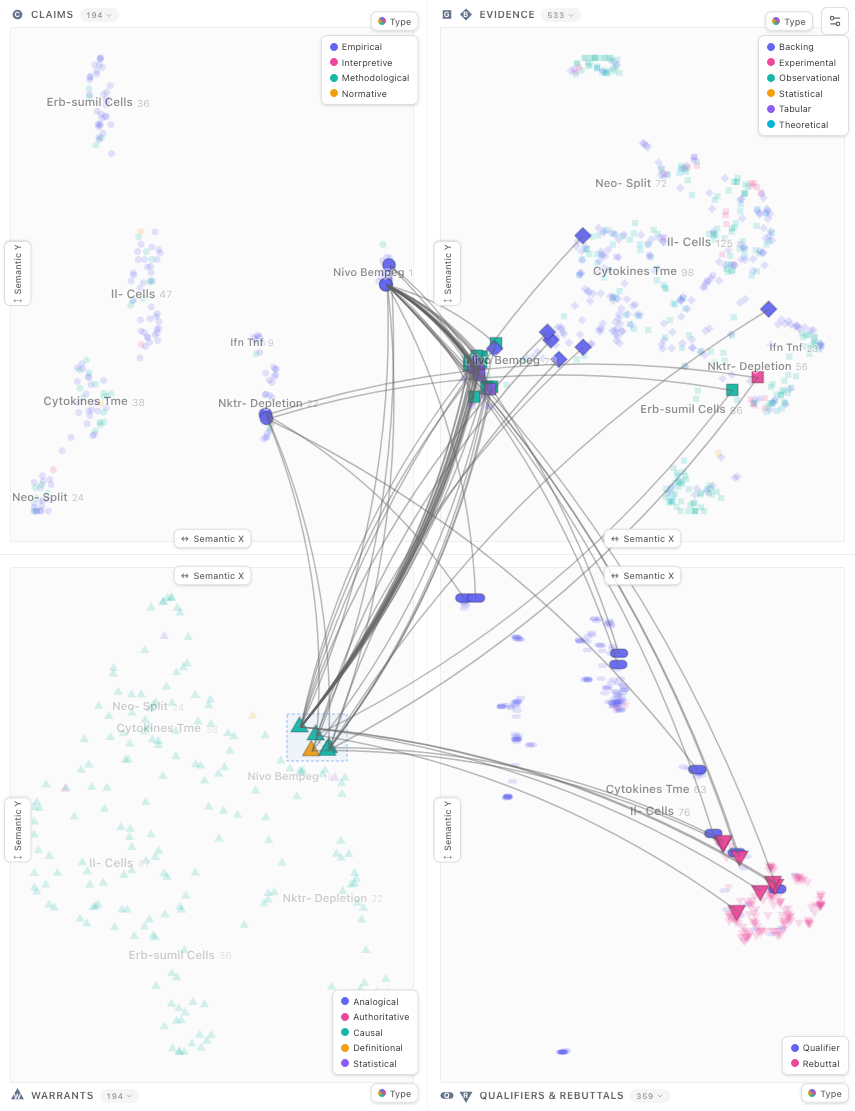}
            \caption{Brushing a region reveals related components across argument roles.}
            \label{fig:corpus-graph}
        \end{subfigure}
    \end{minipage}
    \hfill
    \begin{minipage}[t]{0.52\linewidth}
        \vspace{0pt}

        \begin{subfigure}[t]{\linewidth}
            \centering
            \includegraphics[
                width=\linewidth,
                keepaspectratio
            ]{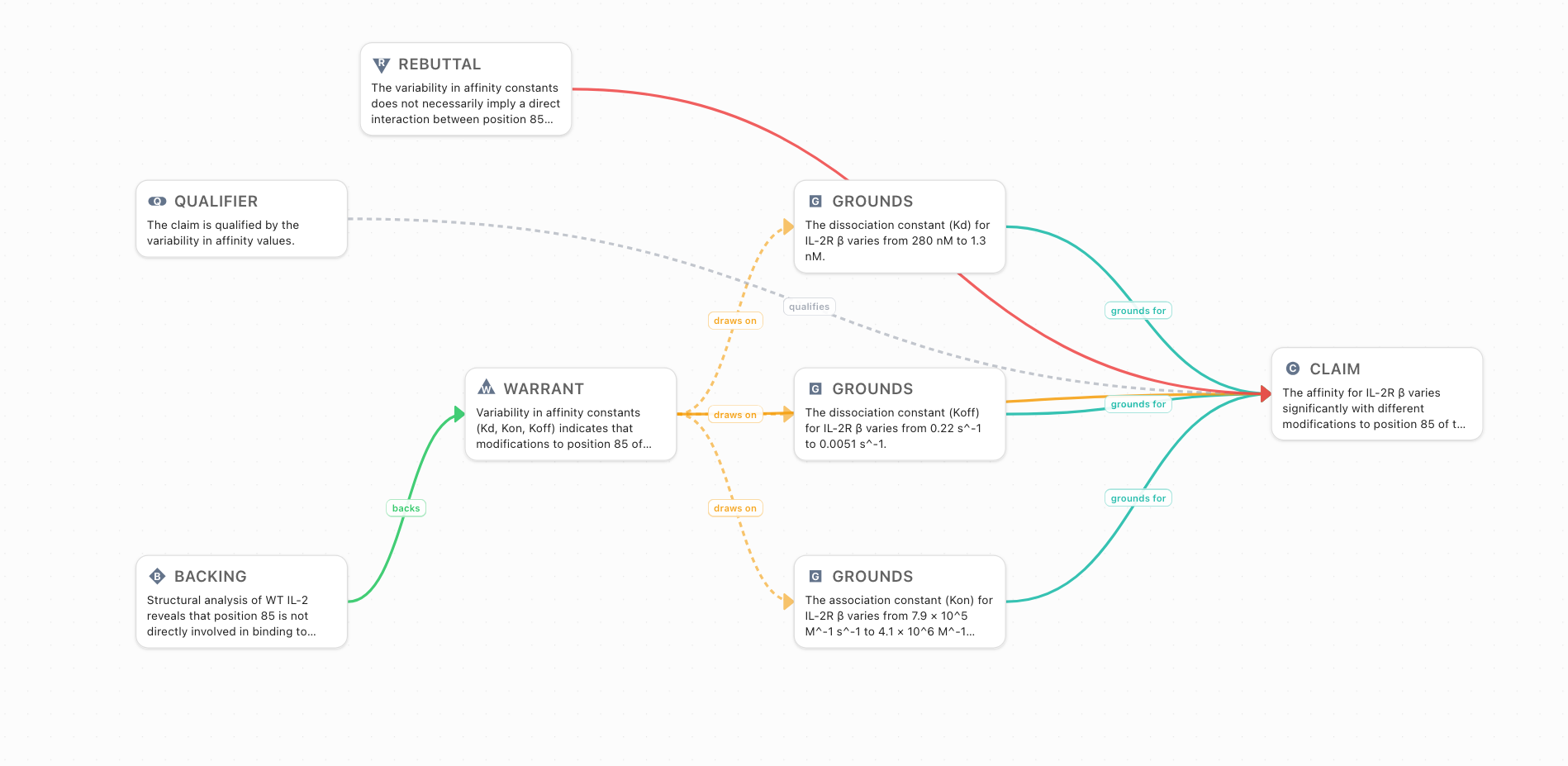}
            \caption{Detailed workbench view of argument components and relations.}
            \label{fig:corpus-detail}
        \end{subfigure}

        \vspace{0.5em}

        \begin{subfigure}[t]{\linewidth}
            \centering
            \includegraphics[
                width=\linewidth,
                keepaspectratio
            ]{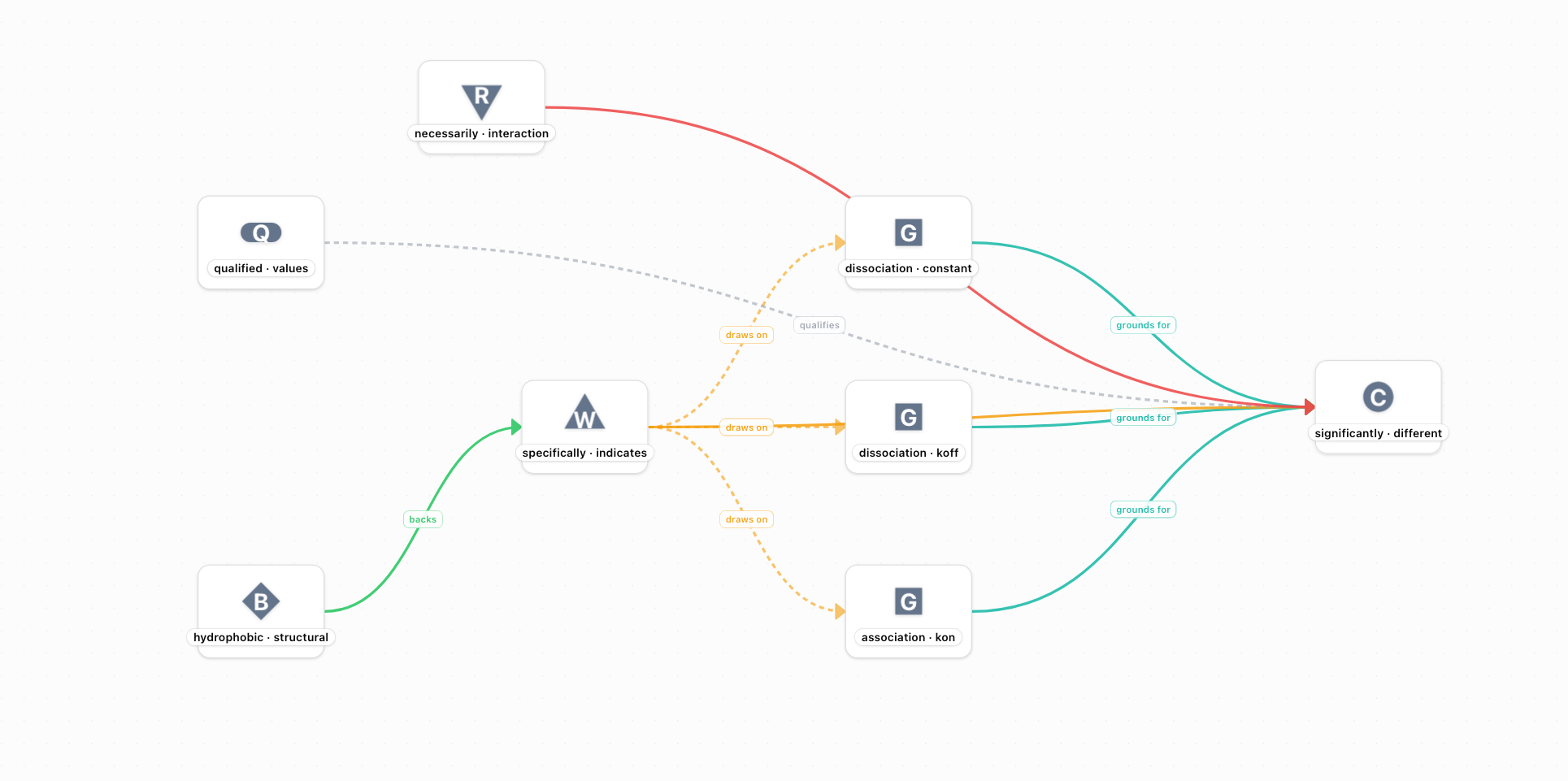}
            \caption{Zoomed-out workbench view with semantic zooming.}
            \label{fig:corpus-overview}
        \end{subfigure}
    \end{minipage}

    \caption{\textbf{Coordinated exploration of argument structure in \toolname.}
    (a) Brushing the argument view exposes relations across role-specific spaces.
    (b) The workbench shows individual argument components and relations at close range.
    (c) Zooming out replaces detailed nodes with abstract shapes to preserve the larger argument structure.}
    \Description{Three views of argument structure in GroundWork. Panel (a) shows four role-specific scatterplots with a brushed region and edges to related components. Panel (b) shows the workbench zoomed in to individual argument components and relations. Panel (c) shows the same workbench zoomed out, where semantic zooming replaces detailed nodes with simplified shapes.}
    \label{fig:corpus_visualization}
\end{figure*}

\subsection{Constructing and Curating Arguments}

% To construct and curate arguments (\textbf{R2}), Analysts can materialize components from the argument view into the workbench.
% Materialized nodes retain a reference to their corpus node and source passage.
% Analysts can also author components directly, including claims to test, new evidence, warrants, and rebuttals.
% Authored nodes are embedded and placed in the corresponding argument space near semantically related corpus nodes, situating new reasoning within the existing corpus.
% The LLM assists the analyst by providing candidate components rather than modifying the inquiry graph directly.
% Analysts can request evidence for a claim, generate claims from selected evidence, identify missing warrants, or propose rebuttals.
% Retrieval considers the current workbench state, selected nodes, provenance, node type, and semantic similarity before returning structured candidates.
% Analysts can inspect and revise these candidates before promoting them into the inquiry graph.
% Once promoted, model-generated components participate in the same relations and operations as corpus-derived and human-authored nodes.
To construct and curate arguments (\textbf{R2}), analysts can materialize components from the argument view into the workbench.
Materialized nodes retain a reference to their corpus node and source passage.
Analysts can also author components directly, including claims to test, new evidence, warrants, and rebuttals.
Authored nodes are embedded and placed in the corresponding argument space near semantically related corpus nodes, situating new reasoning within the existing corpus.
The LLM assists by providing candidate components rather than modifying the inquiry graph directly.
Analysts can request evidence for a claim, generate claims from selected evidence, identify missing warrants, or propose rebuttals.
Retrieval considers the current workbench state, selected nodes, provenance, node type, and semantic similarity before returning structured candidates.
Analysts can inspect and revise these candidates before promoting them into the inquiry graph.
Once promoted, model-generated components participate in the same relations and operations as corpus-derived and human-authored nodes.
For example, an analyst can materialize grounds about reduced cyclist injuries, author a rebuttal about emergency access, and ask the LLM for backing; the proposed safety-guideline node enters the inquiry only after \texttt{PROMOTE}.

\subsection{Comparing Competing Claims}

% To compare competing claims (\textbf{R3}), Analysts place two or more claims in the workbench and start a comparison.
% \toolname retrieves relevant argument components and records how each item bears on each claim as supporting, attacking, or neutral.
% Neutral items are omitted from the active stance view.
% The results are organized as an evidence-by-claim matrix following Analysis of Competing Hypotheses \cite{heuer1999psychology}, with one row per component and claims as columns (Figure~\ref{fig:competing_hypo}).
% Rows are ordered by diagnosticity, prioritizing evidence that distinguishes between competing claims over evidence that bears on them similarly.

\begin{figure}[t]
  \centering
  \includegraphics[width=\linewidth]{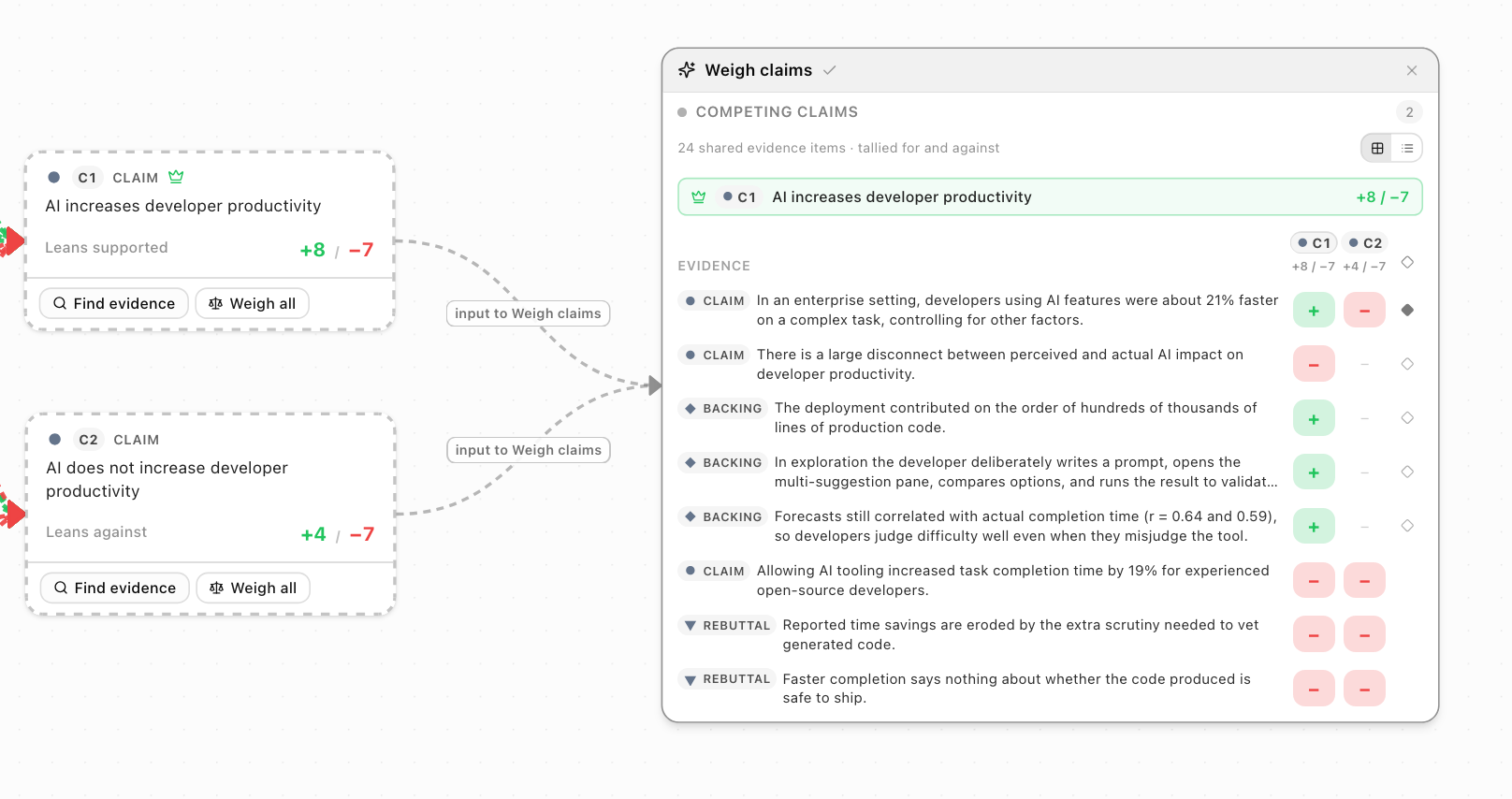}
  \caption{\textbf{Comparing competing claims.}
  Two claims are evaluated against a shared set of argument components, with each matrix cell recording how a component bears on a claim.}
  \Description{Two competing claims in the workbench are connected to a comparison matrix. Rows contain shared argument components, columns represent the claims, and cells indicate whether each component supports, attacks, or is neutral toward each claim.}
  \label{fig:competing_hypo}
\end{figure}

% The matrix exposes where claims rely on shared or conflicting evidence, or different assumptions.
% Analysts can inspect each item's source and argumentative role, compare the warrants connecting evidence to conclusions, and identify which components separate the competing judgments.
% The comparison remains editable throughout the analysis.
% Analysts can add components, revise relations, override model-assigned stances, set evidence aside, or introduce rebuttals, with each change updating the comparison state.
To compare competing claims (\textbf{R3}), analysts place two or more claims in the workbench and start a comparison.
\toolname retrieves relevant argument components and records how each item bears on each claim as supporting, attacking, or neutral.
Neutral items are omitted from the active stance view.
The results are organized as an evidence-by-claim matrix following Analysis of Competing Hypotheses \cite{heuer1999psychology}, with one row per component and claims as columns (Figure~\ref{fig:competing_hypo}).
Rows are ordered by diagnosticity, prioritizing evidence that distinguishes between competing claims over evidence that bears on them similarly.

The matrix exposes where claims rely on shared or conflicting evidence, or different assumptions.
Analysts can inspect each item's source and argumentative role, compare the warrants connecting evidence to conclusions, and identify which components separate the competing judgments.
The comparison remains editable throughout the analysis.
Analysts can add components, revise relations, override model-assigned stances, set evidence aside, or introduce rebuttals, with each change updating the comparison state.
For example, comparing ``add protected lanes citywide'' with ``add them only on busy corridors'' can show that injury reductions support both claims, while an emergency-access rebuttal distinguishes them; setting that rebuttal aside updates the matrix.

\subsection{Testing and Revising Arguments}

% To support argument testing (\textbf{R4}), Analysts can revise the inquiry graph and inspect how its derived judgments change.
% They can add, remove, relabel, or reconnect components, assess them as confirmed, questioned, or rejected, and override or set aside stances from a comparison.
% These actions modify the shared argument state directly rather than producing a separate note or model response.
% Setting evidence aside allows Analysts to test dependencies within an argument.
% A stance item can be temporarily excluded to determine whether support for a claim remains without it.
% The LLM regenerates the derived judgment from the updated inquiry state, exposing dependencies on particular evidence, warrants, or unresolved rebuttals.
% When a state change invalidates a stored model result, \toolname marks that result as stale rather than presenting it as current.
To support argument testing (\textbf{R4}), analysts can revise the inquiry graph and inspect how its derived judgments change.
They can add, remove, relabel, or reconnect components, assess them as confirmed, questioned, or rejected, and override or set aside stances from a comparison.
These actions modify the shared argument state directly rather than producing a separate note or model response.
Setting evidence aside allows analysts to test dependencies within an argument.
A stance item can be temporarily excluded to determine whether support for a claim remains without it.
The LLM regenerates the derived judgment from the updated inquiry state, exposing dependencies on particular evidence, warrants, or unresolved rebuttals.
When a state change invalidates a stored model result, \toolname marks that result as stale rather than presenting it as current.
For example, setting aside the injury-reduction grounds shows whether the citywide bike-lane claim still holds without that evidence; confirming the emergency-access rebuttal can force a narrower qualifier, and any prior model summary is marked stale.

\section{Usage Scenarios} \label{chap:use-cases}

We evaluate \toolname through a design-study approach that combines use-case scenarios with expert feedback \cite{sedlmair2012design}.
% This approach is appropriate for assessing whether a system supports real analytic work, especially when the contribution lies in the fit between a domain problem, a representation, and an interactive artifact.
This section presents scenarios in intelligence, law, and medicine, each centered on a document corpus and a concrete analytic question.
The expert feedback in Section~\ref{sec:expert-feedback} examines whether these workflows, representations, and interactions address recognizable needs in practice.
For all scenarios, corpus graphs are extracted with \emph{Qwen 2.5 14B}, while conversational interactions in the system used \emph{OpenAI's GPT-5.1-nano}.

\subsection{Intelligence: Comparing Competing Judgments}

Strategic intelligence assessments often require analysts to compare competing judgments, examine assumptions, and assess evidence against alternatives \cite{primer2009structured,heuer1999psychology}.
We demonstrate this with an analyst assessing whether the United States can close the rare-earth permanent-magnet supply gap within five years and which assumptions support that judgment.
The scenario foregrounds \textbf{R3} by comparing two opposed judgments against the same body of evidence.

% Strategic intelligence assessments often require analysts to compare competing judgments, examine assumptions, and assess evidence against alternatives \cite{primer2009structured,heuer1999psychology}.
% We demonstrate how \toolname supports this task by helping an analyst test whether a document corpus supports a five-year assessment of rare-earth permanent-magnet supply.
% The analytic question is whether the United States can close the rare-earth permanent-magnet supply gap within five years, and which assumptions would make that judgment defensible.
% This scenario foregrounds \textbf{R3}, because the analyst holds two opposed judgments side by side and reads one body of evidence against both at the same time.

\vspace{0.5em}\noindent\textbf{Corpus.}
The analyst uses five public assessments of the rare-earth permanent-magnet supply chain: the USGS Mineral Commodity Summaries \cite{usgs2025rareearths}, two GAO reports \cite{gao2024criticalmaterials,gao2016rareearths}, a RAND study \cite{willis2023criticalmaterials}, and a Department of Energy report on the NdFeB magnet supply chain \cite{smith2022rareearthmagnets}.
These documents broadly agree that the supply chain is vulnerable due to supply concentration, defense demand, limited substitution, and processing bottlenecks, but provide less consistent evidence about whether the gap can be closed within five years.

% \vspace{0.5em}\noindent\textbf{Corpus.}
% The intelligence analyst uploads five public assessments of the rare-earth permanent-magnet supply chain.
% They are the USGS Mineral Commodity Summaries for rare earths \cite{usgs2025rareearths}, two GAO reports on rare earths and critical materials \cite{gao2024criticalmaterials,gao2016rareearths}, a RAND study on critical-material supply-chain policies \cite{willis2023criticalmaterials}, and a Department of Energy report on the NdFeB magnet supply chain \cite{smith2022rareearthmagnets}.
% The corpus fits because the documents agree about the problem and say much less about the timetable.
% They accept that the magnet supply chain is exposed, through supply concentration, defense demand, thin substitution, and processing bottlenecks.
% They are far weaker, and far less consistent, on whether the United States can close the gap within five years.

\vspace{0.5em}\noindent\textbf{Analysis.}
The analyst authors two competing five-year judgments in the workbench.
\begin{itemize}
    \item[\textbf{C1}] The United States \textbf{cannot} close the rare-earth permanent-magnet supply gap within five years.
    \item[\textbf{C2}] The United States \textbf{can} close the rare-earth permanent-magnet supply gap within five years.
\end{itemize}

\begin{figure*}[t]
  \centering
  \includegraphics[width=\linewidth]{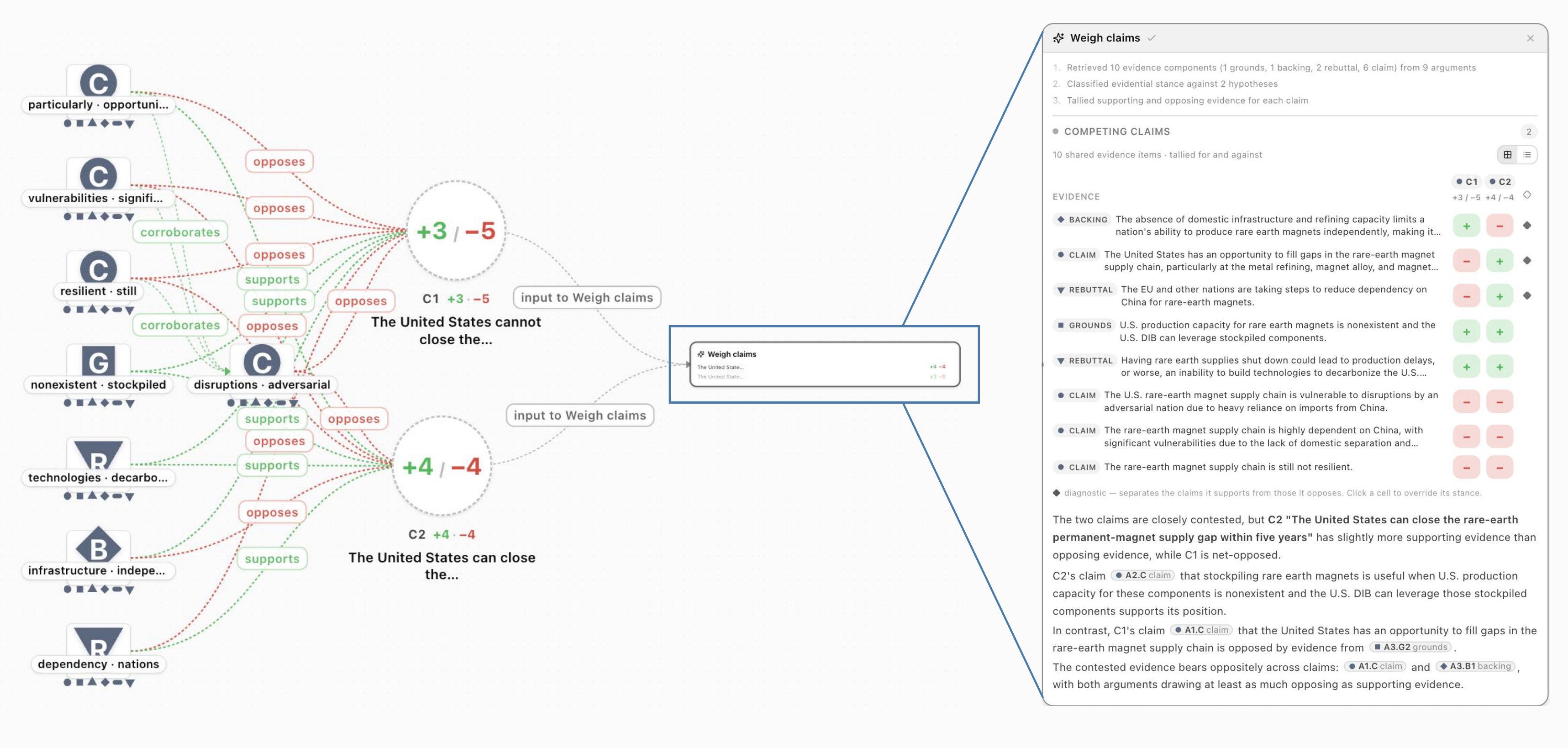}
  \caption{\textbf{\toolname compares two competing claims} about whether the United States can close the rare-earth permanent-magnet supply gap within five years.
  The workbench shows source-linked argument components connected to each claim, while the comparison window lists the retrieved evidence and its stance toward each judgment.}
  \Description{Comparison of two opposing five-year supply judgments. The workbench links source-grounded argument components to both claims, while a comparison table shows each retrieved piece of evidence and whether it supports or challenges each judgment.}
  \label{fig:intelligence-usecase}
\end{figure*}

The analyst then evaluates both judgments against the corpus.
\toolname retrieves evidence relevant to either claim, assigns each piece a stance toward both judgments, and presents the results as a matrix with one row per piece (Figure~\ref{fig:intelligence-usecase}).
Neither judgment is well supported, and the more useful finding is that the evidence does not separate them.
Much of the evidence instead describes conditions shared by both sides, including limited domestic magnet production, reliance on China, constrained separation and refining capacity, and opportunities in refining, alloying, and magnet manufacturing.
Because these conditions underlie both judgments, apparent support for one side often reflects the shared problem rather than evidence that distinguishes the two outcomes.

The analyst treats this lack of separation as the main finding.
Evidence of current vulnerability does not establish whether the gap will close within five years, while identified mitigation opportunities do not show that new capacity can be built, qualified, and scaled within that period.
Reading the matrix alongside the underlying sources reveals that the judgment ultimately depends on whether domestic or allied processing and magnet manufacturing can scale fast enough to reduce dependence on China.
The analyst records these milestones as indicators to watch and adds a qualifier that treats stockpiling as a short-term measure rather than a solution.

\vspace{0.5em}\noindent\textbf{Takeaway.}
A conventional summary would report that the corpus contains both vulnerability findings and mitigation options.
The comparison shows why that is not enough to support a five-year judgment.
The corpus supports a general assessment of exposure and does not support a stable conclusion about closing the gap in time.
What the analyst receives is a structure that can be monitored, holding the two judgments, the evidence that establishes exposure, the options that partly address it, and the milestones that would change the answer.

\subsection{Law: Testing a Causal Argument}

\begin{figure*}[ht]
    \centering

    \begin{subfigure}[t]{0.9\textwidth}
        \centering
        \includegraphics[width=\linewidth]{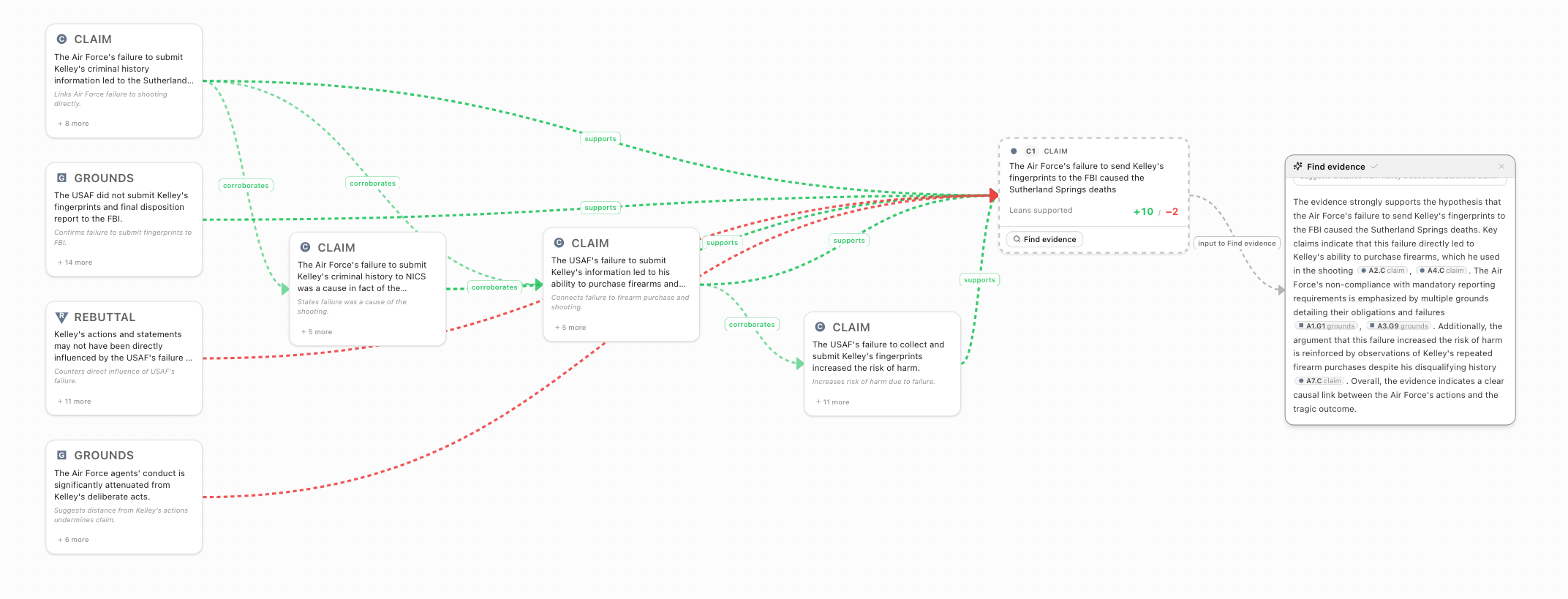}
        \caption{The assembled argument shows that most retrieved evidence supports the families' causal claim, while only a few pieces challenge it.}
        \label{fig:legal-usecase-a}
    \end{subfigure}

    \vspace{0.5em}

    \begin{subfigure}[t]{0.9\textwidth}
        \centering
        \includegraphics[width=\linewidth]{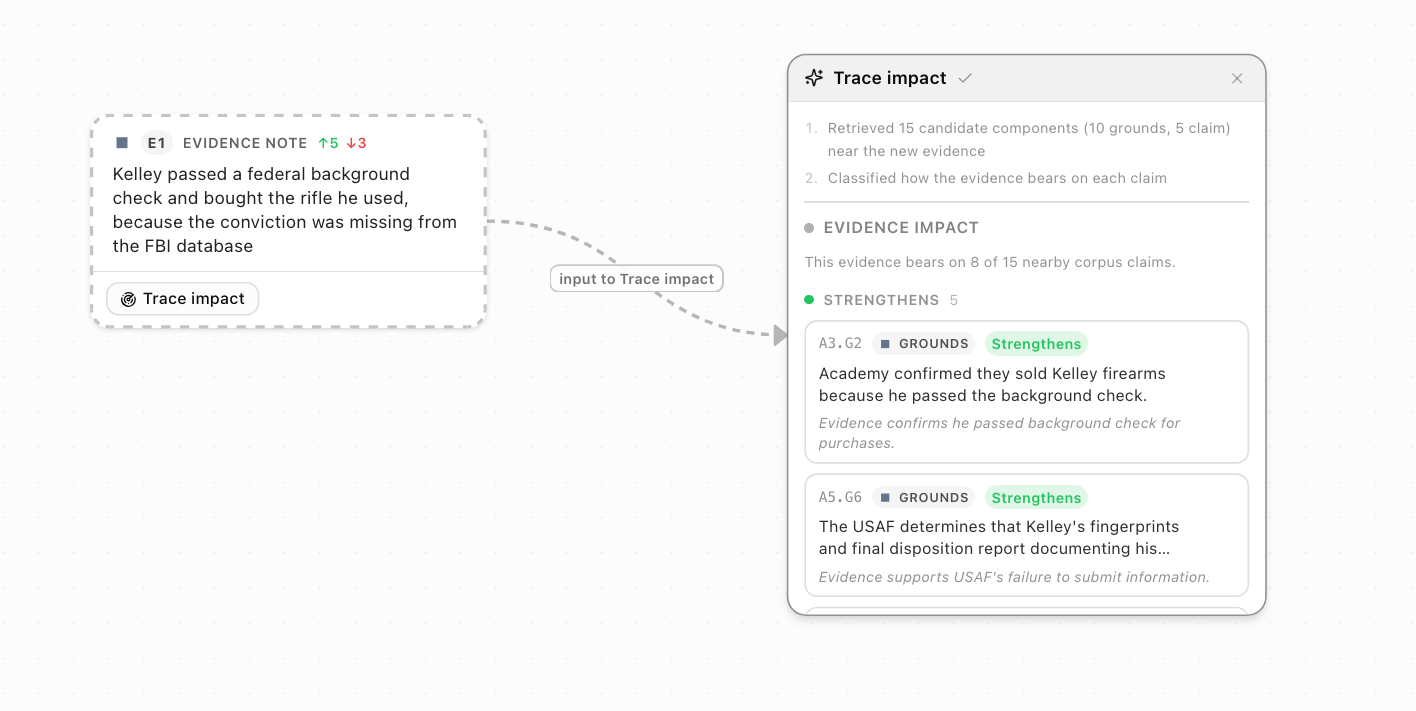}
        \caption{Tracing the added background-check fact shows that it reinforces several existing components, revealing where the argument gains additional support.}
        \label{fig:legal-usecase-b}
    \end{subfigure}

    \caption{\textbf{Legal use-case scenario.}
    The lawyer constructs and tests the causal argument by curating evidence, adding missing components, and revising the argument state.}
    \Description{Two stages of the legal analysis. Panel (a) shows a causal claim connected to source-grounded evidence, with most components supporting the claim and a smaller number challenging it. Panel (b) shows the lawyer tracing a newly added background-check fact to the canvas that it then challenges.}
    \label{fig:legal-usecase}
\end{figure*}

Legal analysis often requires practitioners to connect facts to governing rules, test competing interpretations, and construct defensible conclusions \cite{neumann2005legal,walton2005argumentation}.
We demonstrate this with a fictive Air Force lawyer defending against claims arising from the Sutherland Springs shooting\footnote{\emph{Holcombe v. United States}, 516 F. Supp. 3d 660 (W.D. Tex. 2021).}.
The lawyer asks whether the Air Force's failure to report Devin Kelley's criminal record to the FBI is sufficiently connected to the later shooting to establish causation.
The scenario foregrounds \textbf{R2} and \textbf{R4} as the lawyer curates the retrieved argument, adds missing components, and tests the claim by revising its structure.

\vspace{0.5em}\noindent\textbf{Corpus.}
The lawyer uses six public documents on the reporting failure: the trial court's findings \cite{holcombe2021findings}, the government's appeal brief \cite{usa2023holcombebrief}, an inspector general investigation \cite{dodig2018kelley}, two reporting rules \cite{dodi5505.11,dafi71-101v1}, and a law review note on government liability \cite{note2024discretionary}.
Together, they place findings of fact, party arguments, reporting duties, and legal conclusions in the same corpus.

\vspace{0.5em}\noindent\textbf{Analysis.}
The lawyer begins with the families' claim that the Air Force's failure to report Kelley's fingerprints caused the deaths.
She asks \toolname to gather relevant evidence, which is placed around the claim and is predominantly supportive (Figure~\ref{fig:legal-usecase}a).
The resulting structure captures the expected causal chain: the record was not filed, the background-check system therefore contained no disqualifying record, and Kelley was able to purchase firearms.
This chain is assembled across the trial court's findings, the government's brief, and the inspector general report.

% Two supporting pieces are redundant, so the lawyer sets one aside.
Only two pieces oppose the claim, both attributing responsibility to Kelley rather than the reporting failure.
She marks grounds she verified in the record as confirmed and an unverified inference as questioned.
The support count updates immediately from the revised graph state, while the existing summary is marked out of date.

The lawyer then adds two missing components herself.
First, she adds that Kelley passed a federal background check and purchased his rifle because the conviction was absent from the FBI database.
\toolname identifies this fact as strengthening eight existing pieces and weakening none, including evidence that the sale proceeded because the background check returned clean (Figure~\ref{fig:legal-usecase}b).
She then adds the defense's rebuttal that Kelley planned and carried out the shooting himself, five years after his discharge and six hundred miles from the original investigation.
This joins the two objections already retrieved from the corpus.

Finally, she sets aside evidence from the inspector general report to test how the argument changes without relying on that source.
Half of the supporting evidence disappears, leaving support primarily from the trial court's findings and the government's brief.
The resulting structure shows which parts of the causal argument depend on the inspector general report and which remain supported without it.

\vspace{0.5em}\noindent\textbf{Takeaway.}
A plain summary of these six documents could report the reporting failure, the trial court's finding, and the government's answer.
\toolname instead shows how that material bears on one claim the lawyer keeps editing.
She builds the case against her client, keeps the objections the corpus already held, writes in the fact she cannot avoid and the answer to it, and keeps only what survives her own reading.
What she leaves with is not a case summary but a readable chain, from the duty to report, to the failure to report, to the missing background check, and to the causal step that ties them together.

\subsection{Medicine: Tracing Translational Assumptions}

Translational review often requires researchers to judge whether preclinical findings hold in human patients, examine the assumptions that carry a result across species, and revise claims whose scope proves narrower than stated \cite{van2010can,pound2018possible}.
% We demonstrate this with a researcher assessing engineered interleukin-2 (IL-2) cancer therapies, redesigned over the past decade to preserve the cytokine's anti-tumor activity while reducing the toxicity that limits its clinical use.
We demonstrate this with a researcher assessing a redesigned cancer immunotherapy that showed promising results in animal models but later failed in a clinical trial (Figure~\ref{fig:teaser}).
The researcher seeks to validate the claim that the redesign improves outcomes in patients rather than only in animal models.
The scenario foregrounds \textbf{R1} and \textbf{R5} as the researcher moves from a corpus-level pattern to a single argument, traces evidence to its source passages, and confronts a decisive inferential step that no source states.

\vspace{0.5em}\noindent\textbf{Corpus.}
The researcher uses six papers on engineered interleukin-2 therapies: four preclinical studies in mice, each testing a different redesign and reporting retained anti-tumor activity with reduced toxicity \cite{levin2012superkine,sun2019nextgen,sharma2020bempegaldesleukin,quijanorubio2023split}; the phase III trial of one such redesign, bempegaldesleukin, in advanced melanoma, which did not meet its endpoints \cite{diab2023pivotio001}; and a 2025 review of the field \cite{salazar2025cytokine}.
Together, they place preclinical results and clinical outcomes in the same corpus, where the underlying biology is largely agreed, and the inference from model to patient is not.

\vspace{0.5em}\noindent\textbf{Analysis.}
The researcher authors the claim that the redesigned therapy improves patient outcomes when added to an existing immunotherapy.
She asks \toolname to gather relevant evidence, which is placed around the claim in the workbench (Figure~\ref{fig:teaser}C) and is predominantly supportive, with sixteen supporting pieces against four opposing.
Binding the substrate color channel to the source document reveals that twelve of the sixteen supporting pieces come from the four preclinical studies.
The same split is visible in the argument spaces (Figure~\ref{fig:teaser}B), where preclinical grounds cluster apart from trial outcomes.
The one trial piece among them states the drug's design rationale rather than an outcome, so the apparent support rests almost entirely on animal evidence.

She expands the argument behind the trial's opposing piece, which places the study's own conclusion on the canvas: that the combination provides no additional clinical benefit over the existing therapy alone.
% Its grounds arrive with it and trace to the reported figures: an objective response rate of 27.7\% against 36.0\%, progression-free survival of 4.17 months against 4.99, and grade 3-4 adverse events rising from 11.5\% to 21.7\%.
% Read by source rather than by count, the evidence separates into two populations the claim had conflated.
Its grounds arrive with it and are traced in the source panel (Figure~\ref{fig:teaser}A) to the reported figures: an objective response rate of 27.7\% against 36.0\%, progression-free survival of 4.17 months against 4.99, and grade 3--4 adverse events rising from 11.5\% to 21.7\%.
Read by source rather than by count, the evidence separates into two populations the claim had conflated.

The researcher then scopes the evidence to the claim.
Two opposing pieces describe the toxicity of the original treatment, which motivated the redesign rather than providing evidence about the redesigned therapy itself.
She marks each preclinical result as questioned for applicability rather than removing it, since the finding stands and only its reach is in doubt.
The support count updates immediately from the revised graph state.

The decisive step is the warrant connecting positive results in mouse models to expected benefit in human patients.
No source states this, so there is no passage to inspect, and \toolname marks the warrant as inferred.
The same preclinical paper supplies a rebuttal cautioning that its results may not transfer to human patients because of species differences, which she promotes so that it stands against the claim.
She then authors a narrower claim that the redesign reduces toxicity while preserving anti-tumor activity in mouse models.
The corpus supports the narrow claim and does not support the broad one, and she qualifies the surviving claim to animal models.

\vspace{0.5em}\noindent\textbf{Takeaway.}
A plain summary of these six papers would report positive results in mice and a failed clinical trial.
The argument graph instead shows where those two findings meet, at a single inferential step that no paper states.
That step becomes an object the researcher can select, assess, and set aside, re-counting the claim from the revised graph state.
What she leaves with is not a literature summary but a claim whose scope is marked, together with a record of which evidence ceased to count and why.

% \begin{figure*}[t]
%   \centering
%   \includegraphics[width=\linewidth]{figs/medical.pdf}
%   \caption{\textbf{Use-case scenario in medical decision support.}
%   A reviewer asks whether low-dose aspirin offers net benefit for primary prevention in older adults.
%   \toolname renders the answer not as a verdict but as a contestable tally over source-grounded benefit and bleeding-harm evidence. 
%   Correcting one mislabelled stance and setting aside evidence from younger patients re-tallies the claim instantly and flips it from mixed to refuted for this population.
%   The recommendation is thus not globally supported or unsupported: it holds only within a scope its qualifiers and rebuttals make explicit, which the analyst can inspect and revise rather than trust.}
%   \label{fig:medical-usecase}
% \end{figure*}
\section{Expert Feedback} \label{sec:expert-feedback}

We complement the use-cases in Section~\ref{chap:use-cases} with expert feedback to examine how participants currently construct and evaluate arguments, how they perceive \toolname, and where its structured argument model aligns with or departs from their existing workflows.

% \adam{This goal sounds like a usability study, which this is not. It's expert feedback on their domain workflows. I would go back and reread the Design Study paper to better understand the methodology, and revise this goal statement to match.}

\subsection{Study Design}
\label{sec:study-design}

We conducted semi-structured feedback sessions with four domain experts.
Participants were recruited through the authors' professional networks and are reported in Table~\ref{tab:expert-participants} using broad professional and institutional descriptors to preserve anonymity.

\begin{table}[h]
    \centering
    \small
    \begin{tabular}{llll}
    \toprule
    \textbf{ID} & \textbf{Domain} & \textbf{Credential} & \textbf{Anonymized affiliation} \\
    \midrule
    E1 & Consultancy & Data analyst & Anonymous consultancy company \\
    E2 & Product & Business executive & Anonymous start-up \\
    E3 & Legal & Practicing attorney & Anonymous law firm \\
    E4 & Immunology & Researcher & Anonymous medical school \\
    \bottomrule
    \end{tabular}
    \caption{\textbf{Expert participants.}
    Participants are identified by anonymized expert IDs used for quote attribution in the findings.}
    \label{tab:expert-participants}
\end{table}

Each session consisted of two parts.
First, the interviewer asked a set of questions about the participant's current workflow, including how they locate, organize, evaluate, and retain information across documents (Appendix~\ref{app:study_materials}).
Second, we conducted a guided walkthrough of \toolname using a representative example corpus.
During the walkthrough, participants related the demonstrated interactions to their own work, described what they would inspect or do next, and identified where the representation did not match their domain practice.
We demonstrated the core capabilities of \toolname.
Sessions lasted approximately 60 minutes, were conducted over video conference, and were recorded and transcribed for analysis.

We analyzed the transcripts using a predominantly inductive thematic analysis following Braun and Clarke~\cite{braun2006using}.
After each session, we recorded observations about participants' reasoning practices and reactions to the structured argument representation.
We compared these observations across participants to identify candidate themes, then returned to the transcripts to refine overlapping themes and verify their support in the data.
The research team iteratively discussed and refined the resulting themes.

Following Sedlmair et al.'s emphasis on iteratively refining problem and task abstractions through engagement with domain experts~\cite{sedlmair2012design}, we use these themes to refine our characterization of expert reasoning rather than treating participants' comments as a list of usability problems or feature requests.
Specifically, the themes identify which aspects of reasoning become explicit when argument structure is represented persistently in \toolname, and which forms of professional judgment remain implicit or require additional domain-specific representation.
We return to these refinements in Section~\ref{sec:discussion} to derive broader implications for the design of structured human-LLM argumentation.

\subsection{Study Results}
\label{sec:study-results}

Across the sessions, participants recognized the reasoning represented in \toolname as relevant to their own work.
They related its claims, evidence, assumptions, and counterarguments to practices they already used when working across documents, including note-taking, source tracking, comparison, and maintaining relationships mentally.
Participants also saw high value in keeping these relationships persistent and traceable to the source, particularly when an analysis needed to be revisited or revised over time.

% These reactions suggest that the representation was sufficiently aligned with participants' existing reasoning practices to serve as a basis for reflecting on how that reasoning is actually constructed and evaluated.
These reactions show that the representation aligned with participants' existing reasoning practices closely enough for them to examine how that reasoning is constructed and evaluated.
The more substantive findings emerged from where participants' practices went beyond the structure represented in \toolname.
We organized these findings from across the sessions into three themes.

\vspace{0.5em}\noindent\textbf{Theme 1: Common argument structure, different reasoning processes.} 
% \remco{awkward to read}
Across domains, participants recognized similar argumentative objects in their work, but did not describe a common procedure for constructing them.
For example, E4 mapped scientific reasoning directly onto the Toulmin roles, explaining that \emph{``all the arguments that are made in a paper basically follow this model, and that you can see every one of these node types in immunology as well.''}
% \adam{"In contrast, Ex interpreted argumentative objects in this way: "..." -- provide a counterexample!}
In contrast, E2 interpreted backing as \emph{``my knowledge base, here's stuff that I know,''} warrants as \emph{``inferences from that \ldots{} knowledge base,''} and grounds as \emph{``a sort of stimulus \ldots{} a conversation, a new piece of information, \ldots{} a question.''}
% Groundwork's instantiation of the Toulmin model therefore achieved its goal of making argumentation interactive and inspectable (\textbf{R1-R2}).
% But this may also suggest that the sequence of reasoning over argumentative objects may be less understood, while being a core issue with how people use LLMs to help them construct arguments.
Together, these accounts suggest that Toulmin provided a common structure for making argumentation interactive and inspectable (\textbf{R1-R2}), while the reasoning process around that structure remained more variable.
This variation appeared in three places: how participants moved through the argument, how they kept track of it, and how they divided work with the LLM.

% \adam{was this a design requirement? which R is this?}.
% \adam{I see each paragraph here as demonstrating an observation which helps explain your theme's thesis. they are examples to explain your point.}

% First, we observed the paths participants took through those objects differ substantially.
First, participants followed substantially different paths through the same argumentative objects.
% \adam{, affected by the order that LLMs presented information? -- how does this tie into LLMs?}.
E3 described legal work as claim-first, stating that \emph{``usually, when we deal with cases, we get the claim first''} and then search the available legal material.
E1 instead described an incremental process in which \emph{``you actually know very little''} at the beginning and \emph{``more and more puzzle pieces are added, and you gradually get a clearer picture.''}
E2 described both directions within the same practice, distinguishing \emph{``almost like a proactive versus a reactive version''} depending on the starting point.
These accounts separate the representation of an argument from the procedure used to reason with it.
The same reasoning state can be approached from different entry points and assembled in different orders.
For human-LLM interaction, this suggests separating the structure of the shared reasoning state from the interaction path through which it is constructed, so users can direct the sequence while the model assists.
% \adam{is this the takeaway? what was unclear here? is this new information?}

Second, participants' existing practices also showed that this intermediate state matters beyond the moment in which it is created.
E3 described relationships among evidence and claims as something lawyers ordinarily maintain \emph{``in our brains''}, while E1 collected fragments in OneNote so they could be connected and revisited later.
E4 similarly emphasized preserving an analysis so that \emph{``six months later \ldots{} you could very easily look at that document and remember what the paper was about and what was important.''}
E1 contrasted this evolving state with conversational interaction directly, observing that \emph{``a chat is static, whereas \toolname is dynamic.''}
% Together, these accounts suggest that persistence supports continuity as evidence, interpretations, and source collections change.
% \adam{so the point is that being able to revisit information was critical, but LLMs make it hard to do so when presenting information linearly? somehow draw that out more}
These accounts suggest that persistence matters, but argumentation is not always organized through an explicit schema.
Participants used notes, sketches, and mental connections as flexible ways to keep track of an evolving analysis.
For human-LLM interaction, this might suggest separating the formal structure used by the model from the representation through which users organize their own reasoning.

Third, the division of labor between human and AI was flexible.
Participants were willing to delegate retrieval, organization, summarization, and the construction of an initial structure, but drew a boundary around professional judgment.
E2 described the desired role of AI as \emph{``a lot of remembering, but not a lot of thinking''} and stated that \emph{``decision making is still 100\% something that I do.''}
E1 similarly cautioned that handing over the process could mean that \emph{``you lose some of that control.''}
% Instead, it provided a common state around which both the path of reasoning and the contribution of the model could vary.
% \adam{same thing. What's the role of the LLM? Should it go first, last? On demand or autonomously? If we are saying that we succeeding in making a useful model, but that how people/LLMs use the model is still unclear, then draw out where that tension was.}
Together, these accounts do not point to a single point at which the LLM should enter the workflow.
The shared representation therefore did not imply a fixed workflow or a fixed allocation of agency.
This suggests that the timing and initiative of model contributions should follow the user's reasoning process, rather than define it.

\vspace{0.5em}\noindent\textbf{Theme 2: Argumentative role did not determine evidential standing.}
Making support and opposition explicit surfaced a second judgment that could not be determined by argument structure alone, such as in any automated system.
Participants were responsible for distinguishing between what role a piece of evidence played in an argument and how much that evidence should matter.
Two sources could support the same claim while differing substantially in authority, credibility, recency, methodological status, or contextual applicability.
In this sense, \toolname addressed \textbf{R3} by making how evidence bears on competing claims explicit, while \textbf{R5} exposed the source context needed to judge that evidence.
% \adam{"Groundwork therefore achieved the goal of ..." -- add R}
We also found that the role of the LLM in this process was unclear, but a few examples shed light on how humans and LLMs could co-construct such evidential weights, leveraging the strengths of LLM automation to compliment human judgement and expertise.

For example, the criteria used to establish evidential weight differed across domains.
In law, E3 emphasized institutional authority and recency, explaining that \emph{``we have to consider how recent were the cases. What's the level it's coming from? Is it Supreme Court? Like, is it the Supreme Court of the United States? Then that's gonna take precedence.''}
The hierarchy could override the amount of supporting material itself.
As E3 explained, \emph{``obviously the Supreme Court of the state or the United States, that's gonna take more weight and credibility than the lower tiered.''}
In immunology, E4 instead emphasized methodological status, noting that some papers \emph{``have not actually been peer reviewed yet, and therefore are not really considered established evidence that can contribute to the field.''}
E1 described another dimension of standing in consultancy, where temporal context affected whether information remained applicable.
As E1 explained, \emph{``a piece of text might be from 2026, while the diagram is from 2019, and I have to connect the two. So time is definitely important.''}
These results may suggest an important task for automated systems like LLMs to help surface or track domain requirements separately from argumentation structure, allowing humans the ability to offload information to the system so they can focus on judgment.

These accounts also changed how participants described provenance, an important part of the argumentation process.
Tracing information to a source helped verify that a model had reproduced it correctly.
But we observed sources going beyond that role -- the source itself often provided information needed to judge how the evidence should be interpreted and how much it should contribute to a conclusion.
E3 described this as a routine part of legal reasoning, stating that \emph{``the biggest thing that we do in the background, in our heads, is what's the source? Where is it coming from? How credible is it? That's what we consider every single time.''}
E4 similarly explained, \emph{``I always look at the source. \ldots{} If it says, for example, `Paper X observed or found Y,' then I always go back to the paper itself to check whether that is actually correct.''}
E2 described a related practice with AI-generated summaries as \emph{``trust but verify''}, relying on \emph{``spot checking, especially if it's an area that I do have a lot of familiarity with.''}
It may not be enough for LLMs to simply execute search, but to draw out competing perspectives discovered automatically, on behalf of the user.

% \adam{I'm not understanding this summary paragraph. What did we learn? I tried to draw out some findings and potential interpretations above. Maybe summarize them here?}
% The findings therefore separate three aspects of evidence that can otherwise be collapsed together.
% Argument structure captures how information bears on a claim.
% Provenance exposes where that information came from and the context needed to assess it.
% Evidential standing reflects the expert judgment of how much that information should matter.
% Importantly, the interviews suggest that standing is not a single generic property, since different domains relied on different criteria to establish it.

Taken together, these findings show that evidential assessment cannot necessarily be reduced to support and attack relations or source links alone.
Experts rely on properties such as authority, recency, methodological status, and context to determine how evidence should matter.
This suggests a division of labor in which LLMs surface and organize these properties, including conflicting evidence, while experts retain responsibility for judging their significance.
Argumentative role, provenance, and evidential standing should therefore remain distinct in the representation.

\vspace{0.5em}\noindent\textbf{Theme 3. Explicit structure changed where sensemaking happened.}
% \adam{I don't get this title. This section is more like "Explicit structure abstracted away parts of the learning process." -- as in, the process people use to inspect documents manually actually provides tangible benefits to how they compose arguments. but streamlining it lost those points of growth. Terrance Tao, the famous mathematician, has gone on record describing this as an important issue in doing AI-assisted science and math today. Scientists and mathematicians lose the process that makes for seredipitous, great science to emerge. In other words, we learn things by being in the trenches that makes us better. Are we seeing a similar tension here?}
Decomposing an argument changed what participants could refer to when describing interaction with the model.
Rather than responding to an entire generated answer, participants could identify a particular claim, piece of evidence, or relationship as the object they wanted to inspect or change directly.
This evokes similar advantages to the paradigm of direct manipulation \cite{shneiderman1983direct}, applied to schematizing information.
For example, E2 described the value of breaking \emph{``statements, hypotheses down into individual parts''} so that each could be judged separately, for example \emph{``yeah, that scans, I agree with that''} or \emph{``there's something that's wrong here.''}
E4 made a similar observation, explaining that the structure provides \emph{``a kind of framework that the model can also rely on''} and makes it possible to ask specifically about \emph{``the grounds or specifically about the claims.''}

Yet the same explicitness also placed boundaries around how the reasoning process evolved.
E1 described their existing sensemaking process as part of how they came to understand the material itself.
As E1 explained, \emph{``this is my process. It is my way of working. It is also my way of learning to understand those documents.''}
Moving that process entirely into a predefined model raised a concern that \emph{``you are handing that process over. So then you lose some of that control.''}
E3 expressed the same tension from a different perspective, noting that \emph{``attorneys work in different ways''} and value \emph{``the autonomy to play around with things, instead of having a model \ldots{} do certain things in a rigid way.''}
These results may suggest that LLMs too easily abstract away an important learning process, which provides important training to help craft effective argumentation. Replacing this process with pre-determined argumentation structures may cause experts to become narrow-sighted.

These boundaries concerned more than preferred workflow.
Experts also relied on information whose significance depended on domain context.
E4 emphasized that \emph{``in immunology, a lot of things are very context-dependent''} and repeatedly returned to the need to know which paper each component came from before interpreting relationships across papers.
What counts as a useful decomposition therefore depends not only on the argument itself, but also on which contextual distinctions must remain available to the expert.

Together, these accounts reveal a tradeoff inherent to structured interaction.
Making reasoning explicit gives humans and models named objects on which they can act independently.
At the same time, the representation determines which distinctions become directly addressable and which remain outside the structure.
The challenge is therefore not simply to add more structure, but to decide which parts of reasoning should become first-class interaction objects while preserving ways for experts to work beyond them.

\section{Discussion}
\label{sec:discussion}

Taken together, our findings show that a shared argument representation captures only part of how experts reason with LLMs.
Theme 1 showed that participants recognized common argumentative objects while differing in how they constructed, maintained, and divided work around them.
Theme 2 showed that making argumentative relations and provenance explicit still left evidential standing to expert judgment.
Theme 3 showed a further tension: explicit structure made individual parts of reasoning directly actionable, but constructing and organizing those parts was itself part of participants' sensemaking.

Together, these findings suggest that the design goal should not be to formalize as much of expert reasoning as possible.
Rather, structured human-LLM systems should formalize enough of the reasoning state for humans and models to coordinate, while preserving flexibility in how experts work, judge evidence, and develop understanding.
This shifts the design question from \emph{How much reasoning can be formalized?} to \emph{Which parts of reasoning should be shared with the LLM, and which should remain open to the human?}
We discuss the resulting design implications and costs below.

\subsection{Design Implications for Structured Human-LLM Argumentation}

\vspace{0.5em}
\noindent\textbf{Common reasoning structure without prescribing a common workflow.}
Theme 1 showed that participants recognized common argumentative objects without following a common process for constructing them.
E3 typically began from a claim, E1 described understanding emerging as ``more and more puzzle pieces are added,'' and E2 described both claim-first and evidence-first reasoning.
The same representation can therefore support different paths through an analysis.
Human-LLM systems should allow the model to operate on a persistent reasoning state without requiring analysts to enter, traverse, or construct that state in a fixed order.

The same principle applies to how analysts maintain their own reasoning.
Participants relied on mental connections, notes, and personally organized artifacts rather than a single explicit schema.
A formal representation may therefore be useful for coordinating with an LLM without needing to become the analyst's only representation of the task.
Future systems could maintain structured state underneath while allowing analysts to externalize and organize their reasoning through looser notes, spatial arrangements, annotations, or other personally meaningful forms.

% \vspace{0.5em}
% \noindent\textbf{Use structure to support judgment rather than automate it.}
% Participants welcomed AI support for finding, organizing, and proposing information, but consistently retained responsibility for interpreting evidence and making final judgments.
% Structured representations can support this division of labor by making model contributions explicit objects that analysts can inspect, revise, or reject.
% Future systems should therefore use AI to extend expert reasoning without making model-generated structure an implicit commitment.

\vspace{0.5em}
\noindent\textbf{Let model initiative follow the human reasoning process.}
Theme 1 also showed no single preferred point at which the LLM should enter the analysis.
Participants welcomed support for retrieval, organization, summarization, and identifying gaps, but differed in when that support was useful and consistently retained control over professional judgment.
This suggests that the question is not simply whether an LLM should act autonomously or on demand.
Its initiative should instead depend on the state of the human's reasoning: sometimes constructing an initial structure, sometimes responding to a targeted request, and sometimes remaining in the background while the expert works.
A shared representation can support this flexibility because the model can contribute to the same evolving state without determining the sequence of the analysis.

% \vspace{0.5em}
% \noindent\textbf{Separate argumentative role from evidential standing.}
% Representing evidence as supporting or challenging a claim does not determine how strongly it should influence the resulting judgment.
% Participants additionally considered properties such as authority, credibility, recency, methodological quality, and contextual applicability.
% Structured reasoning systems should therefore represent how evidence relates to an argument separately from judgments about how much that evidence should matter.

\vspace{0.5em}
\noindent\textbf{Use LLMs to prepare evidential judgment, not collapse it.}
Theme 2 showed that an argumentative relation captures only one part of how experts evaluate evidence.
Participants additionally considered authority, recency, methodological status, credibility, and contextual applicability, and the relevant criteria differed by domain.
This suggests a useful role for LLMs beyond retrieving evidence or labeling it as supporting or opposing.
Models can surface competing evidence, recover relevant source properties, and organize the information an expert needs to compare its standing.
The judgment of how much those properties should matter, however, remains an expert task.
Structured systems should therefore keep argumentative role, provenance, and evidential standing distinct rather than combining them into a single model-generated assessment.

% \vspace{0.5em}
% \noindent\textbf{Keep shared structure flexible and domain-legible.}
% Explicit structure can make human-LLM interaction more precise by exposing the relationships relevant to a reasoning task.
% However, participants also relied on domain-specific concepts, artifacts, and personal sensemaking practices that did not always map directly onto a fixed representation.
% Systems can therefore share an underlying reasoning structure while adapting its presentation and interaction to the language and practices of the domain.

\vspace{0.5em}
\noindent\textbf{Preserve the sensemaking that happens while structure is created.}
Theme 3 showed that the work of constructing argument structure can itself contribute to expert sensemaking.
E1 explicitly described working through documents as both a way of organizing information and a way of learning to understand it, while E3 emphasized the value of being able to ``\textit{play around with things}.''
Decomposition made claims, evidence, and relationships easier to address individually, but automatically producing that decomposition can also bypass some of the work through which experts notice context, form connections, and refine their understanding.
Human-LLM systems should therefore distinguish between merely burdensome work and work that contributes to sensemaking.
Rather than always presenting a completed structure, systems might progressively expose, propose, or partially construct it so that analysts remain involved in the reasoning that matters.

\subsection{The Costs of Making Reasoning Explicit}

Making reasoning explicit changes more than how information is displayed.
Once claims, evidence, warrants, and rebuttals become persistent objects, both humans and LLMs can refer to and revise the same reasoning state.
At the same time, formalization requires choices about which distinctions are represented and how they are related.
Our findings suggest that Toulmin provides a useful common structure across domains, while experts still interpret and use that structure through their own practices.

This distinction matters because an explicit argument is not necessarily a correct one.
Theme 2 showed that support relations and provenance can make evidence easier to inspect, while judgments about authority, recency, methodological quality, and contextual fit remain with the expert.
A coherent argument structure should therefore be understood as a surface for examination rather than as evidence that the reasoning itself is sound.
This becomes especially important when LLMs contribute components or relations, since those contributions still require interpretation and assessment by the expert.

Theme 3 adds a second consideration.
Some of the work involved in searching, arranging, and connecting information is also part of how experts develop an understanding of the material.
Automatically producing more of the structure may reduce effort, but it can also shift where that sensemaking takes place.
The design challenge is therefore not to preserve every manual step, but to distinguish work that can be safely offloaded from work through which the expert forms useful connections and judgments.

These trade-offs also help define when persistent structure is most valuable.
For analyses in which claims, evidence, and assumptions must be compared, revised, or revisited over time, maintaining an explicit reasoning state can provide continuity that conversational history alone does not.
For simpler retrieval, summarization, or question answering, the same structure may be unnecessary.
Structured argumentation is therefore best viewed as a complement to natural-language interaction, particularly for reasoning tasks whose state must remain inspectable and revisable as the analysis develops.

\subsection{Limitations and Future Work}
Our evaluation focused on how structured argumentation aligns with and challenges expert reasoning practices, rather than on benchmarking \toolname against existing interfaces.
The expert study was small and formative, with participants responding to guided walkthroughs in one-time sessions.
This design allowed us to examine how experts interpreted the representation and where it aligned with or departed from their own practices, but does not speak to longer-term use or comparative task performance.
Future work could build on these findings through longitudinal or comparative studies that examine how structured representations shape what experts notice, remember, revise, and delegate to an LLM over time.

The current system also leaves several opportunities for extension.
\toolname assumes a bounded corpus and has not yet been designed for substantially larger or continuously changing document collections.
Future systems should investigate representations that accommodate changing evidence, collaboration, and the domain-specific properties experts use to establish evidential standing.
They should also explore interactions that allow the human-facing organization of reasoning to remain flexible while maintaining enough underlying structure for model action.
The current representation also has limited support for conditional reasoning, where a claim holds only if another condition is satisfied.
Future work should examine how such dependencies can be represented explicitly.
Further work is needed to improve coverage of complex document content such as tables and to help analysts identify evidence that may still be missing.

\section{Conclusion}

Argumentation over document corpora requires experts to maintain relationships among claims, evidence, assumptions, qualifiers, and counterarguments that are often distributed across documents and difficult to preserve through conversational interaction alone.
We introduced \toolname, a mixed-initiative system that operationalizes the Toulmin model as a shared, editable argument state for humans and LLMs.
Through this representation, users and models can inspect, trace, compare, and revise the same source-grounded reasoning objects as an analysis develops.
Our evaluation further shows that a common argument structure can support different reasoning paths while leaving evidential standing and expert judgment distinct from the structure itself.
Together, this work shows how structured argumentation can provide a persistent foundation for human-LLM reasoning without prescribing a single workflow or replacing expert judgment.

% \begin{acks}
% \end{acks}

\bibliographystyle{ACM-Reference-Format}
\bibliography{bibfile}

\clearpage
\appendix
\section{Use of Generative AI}

Generative AI tools were used to support writing and editing throughout the preparation of this manuscript, including improving grammar, clarity, consistency, and professional tone.
Generative AI was also used to assist in developing and refining the prompts used for Toulmin argument-graph extraction.
AI coding assistants were used selectively during system development to support implementation tasks, while the system architecture and core software design were developed by the authors.
All research ideas, conceptual contributions, study design, analysis, and conclusions presented in this work are original to the authors.

\section{Illustrative Example of Toulmin representation} \label{app:example_toulmin}

This appendix extends the protected-bike-lane example from Figure~\ref{fig:ir_overview} to illustrate how the argument changes through the operation vocabulary.
The figure shows the argument as a set of typed Toulmin components.
Here, we use the same example to show how those components can be added, assessed, and revised during an inquiry.

For example, a human may decide that the claim is too broad and add a narrower qualifier:

\begin{quote}
\emph{Only on corridors with documented cyclist-injury risk.}
\end{quote}

This is represented by \texttt{AUTHOR(Q,text)} followed by \texttt{LINK(Q,C,qualifies)}.
The human may then inspect the emergency-access rebuttal and mark it as confirmed with \texttt{ASSESS(R,confirmed)}.
If the rebuttal should temporarily be excluded from a comparison, it can be neutralized through \texttt{SET\_ASIDE(s)}.
If the LLM proposes a new backing statement, such as evidence from a transportation safety guideline, the proposal is first stored with \texttt{RECORD(y)}.
It changes the inquiry graph only if the human accepts it through \texttt{PROMOTE($\delta$)}.

The example shows why the operation vocabulary is part of the representation rather than an implementation detail.
Without operations, the graph would only describe the structure of an argument.
With operations, it becomes a reasoning state that can be revised over time.
Each edit changes what is active in the inquiry graph, and derived judgments can be recomputed from the updated state rather than regenerated as a separate text response.

\section{Graph Extraction} \label{sec:argument-graph-extraction}

This appendix documents the extraction pipeline used to construct the corpus graph on which \toolname operates.
The pipeline takes a closed collection of user-uploaded documents and an analytic goal as input, and produces typed Toulmin components linked to their source passages.
The analytic goal provides context for identifying and organizing claims, grounds, warrants, backing, qualifiers, and rebuttals.
% Figure~\ref{fig:extraction_pipeline} summarizes the pipeline.
The complete prompts used at each stage of the extraction pipeline are provided in the supplementary material.

The pipeline first processes each document independently.
Each document is parsed, segmented into discourse units, and passed to an LLM with a structured extraction prompt.
% Appendix \ref{sec:extraction-prompts} describes the different prompts used for the extraction.
For each segment, the model returns typed Toulmin nodes.
Segments without detectable argument structure are discarded.
The extraction prompt distinguishes the role of each component.
% Grounds capture direct evidence for a claim, while backing supports the inferential link between evidence and claim.
% Because such links are often implicit, the model is instructed to make them explicit only when supported by the source context.
A second LLM pass checks extracted warrants against their claims and grounds, producing a status such as confirmed, modified, or rejected, and probes the argument with critical questions, adjusting the qualifier strength accordingly.
A grounding check verifies whether each grounds statement is supported by the surrounding source text and stores the relevant passage.

After document-level extraction, \toolname organizes the graph across the corpus.
All extracted components are embedded to support similarity search.
Claim embeddings drive clustering and cross-document relation detection, while warrant embeddings expose recurring inferential patterns.
The system also identifies edges between claims from different documents (corroboration, replication, refinement, and contention) as well as similarity between the warrants that connect them.
Finally, the extracted components are projected into the four argument spaces used by the interface: claims, grounds and backing, warrants, and qualifiers and rebuttals.

% \begin{figure*}[t]
%   \centering
%   \includegraphics[width=\linewidth]{figs/extraction_pipeline.pdf}
%   \caption{\textbf{Argument-graph extraction pipeline.}
%   \toolname transforms an uploaded document collection into a source-linked Toulmin argument graph.
%   Documents are parsed, segmented, and processed by structured LLM extraction to produce typed argument components, which are then verified, grounded in source passages, organized across the corpus, and projected into semantic spaces for analysis.}
%   \label{fig:extraction_pipeline}
% \end{figure*}

\section{Expert Study Questions} \label{app:study_materials}

This appendix reproduces the interview guide used in the expert study described in Section~\ref{sec:expert-feedback}.
The first part of each session covered participants' professional background, document workflows, current sensemaking practices, and experience with AI tools.
The final part asked participants to respond to five statements on a five-point scale, followed by discussion where relevant.

\subsection*{A. You and Your Work}

\begin{enumerate}
    \item What is your current role or title?
    \item What field, discipline, or area of practice do you work in?
    \item How many years of experience do you have in your field?
    \item How many years have you spent doing this kind of analytical or research work?
    \item What type of organization do you work for?
    \begin{itemize}
        \item[] University or research institute; government or military; legal practice; nonprofit or NGO; industry; student; other.
    \end{itemize}
    \item What kinds of document-based work do you typically do?
    \begin{itemize}
        \item[] Examples: literature review or synthesis; case preparation; policy analysis; argumentation; investigation or fact-finding; technical or scientific review; writing; advising decision-makers; teaching; journalism.
    \end{itemize}
    \item What kinds of questions or projects typically land on your desk?
\end{enumerate}

\subsection*{B. Working with Documents}

\begin{enumerate}
    \item In a typical week, roughly how much time do you spend reading or searching through documents?
    \begin{itemize}
        \item[] Probe if useful: under 5 hours; 5 to 10; 10 to 20; 20 to 30; over 30.
    \end{itemize}
    \item On a substantial piece of work, roughly how many distinct documents or sources are involved?
    \begin{itemize}
        \item[] Probe if useful: under 10; 10 to 25; 25 to 50; 50 to 100; over 100.
    \end{itemize}
    \item What kinds of material do you typically work with?
    \begin{itemize}
        \item[] Examples: research papers; books; case law or regulations; institutional reports; technical reports; datasets; interviews or transcripts; news; standards or policy documents; internal correspondence.
    \end{itemize}
    \item Where does this material usually come from?
    \begin{itemize}
        \item[] Examples: databases, archives, clients, colleagues, subscriptions, or personal collections.
    \end{itemize}
    \item What part of working through a large body of material takes the most time or effort?
\end{enumerate}

\subsection*{C. How You Work Today}

\begin{enumerate}
    \item What tools do you currently use for this work?
    \begin{itemize}
        \item[] Examples: Google Scholar; PubMed or other discipline databases; Westlaw or Lexis; Zotero, Mendeley, or EndNote; NotebookLM; Word or Google Docs; Excel or Sheets; Notion, Obsidian, or OneNote; PDF annotation; paper notes; SharePoint; email; domain-specific tools.
    \end{itemize}
    \item Once you have read the material, how do you organize what you found?
    \begin{itemize}
        \item[] Examples: outline, chronology, memo, spreadsheet, mind map, notes, or mentally.
    \end{itemize}
    \item How do you keep track of information that argues against your position or conclusion?
    \item When you hand work to someone who needs to check it or build on it, what do you give them?
    \item What parts of your current process work well enough that you would not want them changed?
\end{enumerate}

\subsection*{D. Experience with AI and Large Language Models}

\begin{enumerate}
    \item How often do you use AI or LLM-based tools?
    \begin{itemize}
        \item[] Probe if useful: never; tried once or twice; a few times a year; monthly; weekly; daily.
    \end{itemize}
    \item Which AI tools, if any, do you use?
    \begin{itemize}
        \item[] Examples: ChatGPT; Claude; Gemini; Copilot; Perplexity; NotebookLM; Elicit; Consensus; SciSpace; Westlaw or Lexis AI; tools integrated into existing software.
    \end{itemize}
    \item What do you use these tools for?
    \begin{itemize}
        \item[] Examples: drafting; summarization; research; brainstorming; editing; data or code analysis; translation.
    \end{itemize}
    \item Are AI tools permitted in your work?
    \begin{itemize}
        \item[] Probe if useful: yes; yes with restrictions; no; unclear or no known policy.
    \end{itemize}
    \item If you have used AI for this kind of work, has anything gone wrong or made you distrust its output?
\end{enumerate}

\subsection*{E. Statements for Discussion}

Participants were asked to respond to the following statements on a five-point scale from \emph{strongly disagree} to \emph{strongly agree}, followed by discussion where relevant.

\begin{enumerate}
    \item I regularly have to weigh competing interpretations of the same evidence.
    \item I regularly have to check that a source actually supports what it is cited for.
    \item Finding what argues against my position is harder than finding what supports it.
    \item I would rely on an AI tool's summary of a source without reading the source myself.
    \item The hard part of my work is finding the right material, not making sense of what I have found.
\end{enumerate}

\end{document}